\documentclass[12pt]{article}
\usepackage{epsfig}
\begin{document}
\thispagestyle{empty}

\begin{flushleft} \bf

RUSSIAN ACADEMY OF SCIENCES

PETERSBURG NUCLEAR PHYSICS INSTITUTE
\setcounter{page}{0}
\vspace{2cm}
\begin{flushright}
Preprint 2362\\
June 2000
\end{flushright}

\vspace{4cm}
S.~I.~Manayenkov

\vspace{2cm}
{\Large{\bf 
New Method for Data Treating in Polarization Measurements } }

\vfill
Gatchina --- 2000
\end{flushleft}
\newpage
\begin{center}
{\bf Abstract}\\
\end{center}
\parbox[t]{\textwidth}
{
Precise formulas are derived for the expected values $<\xi>$,
$<\eta>$ and  variances
$\delta \xi ^2$, $\delta \eta ^2$ of random variables $\xi$, $\eta$
describing the spin asymmetry of
some reaction when a background process contribution is
negligible and appreciable,
respectively. The variances of $\xi$ and $\eta$ are proved
to be finite. This property
differs from that of the Caushy distribution which has an
infinite variance. It is shown that
$<\xi>$ is equal to the
physical asymmetry which allows to find the asymmetry
from experimental data without studying
the detector efficiency. This is the base of the proposed method of data
treating. Asymptotic formulas for $<\eta>$
and $\delta \eta ^2$ are also derived for a
total number of events tending to infinity  for a finite  
value of the background to signal
ratio.
}
\vspace*{0.3cm}
%%\newpage
%\renewcommand{\thefootnote}{\arabic{footnote}}
%\setcounter{footnote}{4}

%%\newpage
%%\renewcommand{\thefootnote}{\arabic{footnote}}
%%\setcounter{footnote}{0}
\section{Introduction}

The study of the polarization phenomena is of great importance in modern physics 
since it gives a valuable information about the spin structure of interaction 
under investigation. Typical examples of the polarization measurements to 
be discussed in the present paper are the measurements of spin-spin asymmetries 
in inclusive deep inelastic scattering of leptons on nucleons (which give the 
spin-dependent structure functions of nucleons) or spin-spin asymmetries in 
semi-inclusive 
processes. A study of a decaying particle polarization in a final state of some 
reaction can be performed by measuring  spin asymmetries in an angular 
distribution of detected particles. To get a high statistical
accuracy of  measured polarization observables we should integrate the result over all kinematic
 variables of which the studied physical quantity is independent. To perform such 
an integration we have to know  the detector efficiency. Modern detectors are 
very complicated devices and a problem to describe an efficiency of the particle 
registration and the detector acceptance is very non-trivial. Therefore approaches in
which one does not need the study of the detector efficiency have a great 
advantage.

Let us consider some examples which explain our approach. To study the polarization 
of the $ \Lambda ^0 $ hyperon we may investigate the angular distribution of the pions 
and protons in the weak decay
\begin{equation}
\Lambda ^0 \rightarrow p + \pi ^-\;.
\label{lambdecay}
\end{equation}
It is well known that the proton angular distribution in the hyperon rest  system 
looks like
\begin{equation}
W(\vec{S},\vec{P})=\frac{1}{4 \pi} (1+ \alpha S \cos \theta )=
\frac{1}{4 \pi} [1+ \alpha (\vec{S}\vec{P})/P]
\label{ldecdist}
\end{equation}
where $\theta$ is the angle between the proton three-momentum $\vec{P}$ 
 ($P=|\vec{P}|$) and the hyperon polarization vector $\vec{S}$ ($S=|\vec{S}|$),
$\alpha =0.642 \pm 0.013$ \cite{PDG} is the well known constant of the weak decay 
$(\ref{lambdecay})$. The mean value of the detected number of the protons is
\begin{equation}
\Delta n(\vec{K},\vec{P},\vec{S})=N_{\Lambda} E(\vec{K},\vec{P},\vec{R})
W(\vec{S},\vec{P}) \Delta \Omega _P \Delta \Omega _K 
\label{protnum}
\end{equation}
where $N_{\Lambda}$ is the total number of the decaying hyperons, 
$\Delta \Omega _P $ is 
the solid angle of the registered protons and  $\Delta \Omega _K$ denotes the    
solid angle of the decaying hyperon momenta, $E(\vec{K},\vec{P},\vec{R})$
is the detector efficiency. The latter depends on the hyperon momentum $\vec{K}$, a position of 
the $\Lambda ^0$ decay vertex $\vec{R}$ and the proton  momentum $\vec{P}$. 
We would like to remark that we shall not discuss in the present work 
the problem of smearing which can, in principle, be important for the detector 
efficiency. We see
from $(\ref{protnum})$ that we have to know $E(\vec{K},\vec{P},\vec{R})$, 
depending on many variables, to obtain the angular distribution 
$W(\vec{S},\vec{P})$ from the measured
quantity $\Delta n /( \Delta \Omega _P \Delta \Omega _K )$. To reduce the statistical
 uncertainty of the obtained polarization of $\Lambda ^0$ we are to fit 
the angular distribution  $W(\vec{S},\vec{P})$ within the total detector 
acceptance for the proton registration.

If it is possible to inverse the direction of the $\Lambda ^0$ polarization  we may 
study the physical asymmetry
\begin{equation}
C=\frac{\Delta n _+ -\Delta n _-}{\Delta n _+ + \Delta n _-}
\label{physasymm}
\end{equation}
where
\begin{equation}
\nonumber
\Delta n _{\pm}=  \Delta n(\vec{K},\vec{P}, \pm \vec{S})\;.
\end{equation}
It is easy to see from $(\ref{protnum})$, $(\ref{ldecdist})$ and $(\ref{physasymm})$ 
that 
\begin{equation}
C=\alpha S \cos \theta \;.
\label{cscos}
\end{equation}
We see from  $(\ref{cscos})$ that $C$ is proportional to the hyperon polarization and 
does not depend on the detector efficiency if the solid angles $\Delta \Omega _K$
and $\Delta \Omega _P$ are small enough. In this approach we do not need to know 
 the detector efficiency but we have encountered  two problems. The quantities of  
$\Delta n _+$ and $\Delta n _-$ in $(\ref{protnum})$ and $(\ref{physasymm})$
are the  expected values of the observed numbers $p$ and $n$ of decays
$(\ref{lambdecay})$ for the positive ($p$) and negative ($n$) hyperon polarization, 
respectively. If we consider the random variable $\xi$ 
\begin{equation}
\xi=\frac{p-n}{p+n}
\label{randxi}
\end{equation}  
then its  expected value $< \xi > $  can be quite different
from the physical asymmetry $C$ given by  $(\ref{physasymm})$ or by the equivalent 
relation
\begin{equation}
C=\frac{<p> -<n>}{<p> +<n>}
\label{asymm2}
\end{equation} 
for finite (and even small) values of $<p>$ and $<n>$. We denote as $<p>$ and $<n>$ 
in $(\ref{asymm2})$ the  expected values of the random variables $p$ and $n$, 
respectively.
We may suspect that $<\xi> \rightarrow C$ if $\Delta n _- =<n> \rightarrow \infty$,
$\Delta n _+ = <p> \rightarrow \infty$. The former problem  mentioned above is 
whether the difference
$<\xi> -C$ is large or not at finite  values of $<p>$ and $<n>$. It will 
be shown later that $<\xi> = C$ if one may neglect the background 
 contribution.

In the approach under consideration we make use of only a small fraction of experimental
events which belong to the solid angle $\Delta \Omega _P $. The latter problem 
consists in   utilization of all the observed events to reduce the 
statistical uncertainty of the obtained
 polarization of $\Lambda ^0$. This problem can be non-trivial if the probability density
of $\xi$ has the same property as the Cauchy density.
Indeed, the Caushy random variable is equal to the ratio of two random variables
having the Gaussian probability densities. 
It is well known (see for example \cite{Hudson},\cite{Martin}) 
that the Caushy distribution  has an infinite  second moment and 
the probability to observe values of the Cauchy random variable which
deviate significantly from the expected value is large. Moreover, the sample mean
of $N$ Cauchy random variables has the same probability density as every
random variable. This means that one does not reduce the statistical uncertainty 
of the measured quantity considering the sample mean.
 But the random variable $\xi$ is just the ratio of two random variables $p-n$ 
and $p+n$ (see $(\ref{randxi})$).  We shall show that the variance 
$\delta \xi ^2$ 
of $\xi$ is finite if we use the conditional probability for  $\xi$ and
the Poisson densities for the random variables $p$ and 
$n$. But it is just the case for the real experimental numbers of events which are 
positive integers.  
This result allows to use the total number of events to reduce the 
statistical uncertainty of the obtained 
polarization of $\Lambda ^0$. It is well known how to do this.
Indeed, let us divide the total  kinematic region of
$\Omega _P$ into $N$ bins $\Delta \Omega _1$, $\Delta \Omega _2$, ..., $\Delta \Omega _N$
and define the random variables $\xi _j$, $\zeta _j$ in the $j$th bin by the relations
\begin{equation}
\xi _j=\frac{p_j-n_j}{p_j +n_j}\;,
\label{defxijzetj1}
\end{equation}
\begin{equation} 
\zeta _j=\frac{\xi_j}{b _j}\;,
\label{defxijzetj2}
\end{equation} 
\begin{equation}
b _j=\alpha \cos \theta _j\;
\label{defxijzetj3}
\end{equation} 
where $p_j$ and $n_j$ denote the numbers of $\Lambda ^0$ decay events for the 
positive and negative hyperon polarization, respectively.
In $(\ref{defxijzetj3})$ $\theta _j$ is a value of the angle between the $\Lambda ^0$ 
polarization and the proton momentum in $j$th bin. 
 Let us choose $M$ bins ($M \leq N$) from $N$ bins 
and denote them as 1th, 2nd, ..., Mth.   
Since the  expected values of $\xi_j$ are equal to $C$ given by
$(\ref{cscos})$ (we are to replace $\theta$ with $\theta _j$ for the $j$th bin) 
 the random variables $\zeta _j$ for every $j$ and $\zeta $ have the 
 expected values 
$<\zeta _j>$ and $<\zeta >$ equal to the $\Lambda^0$ polarization $S$ where
\begin{eqnarray}
\zeta =\sum _{j=1}^M \beta _j \zeta _j\;
\label{defzet}
\end{eqnarray}  
and the coefficients $\beta _j$ are positive numbers  obeying the relation
\begin{eqnarray}
\sum _{j=1}^M \beta _j =1\;.
\label{sumbet}
\end{eqnarray}  
In a general case $b_j$ are given by  expressions other than $(\ref{defxijzetj3})$  but
they represent some known functions. They are used to make the expected values of all $\zeta _j$
equal to each other.
It is well known \cite{Hudson} that $ \beta _j$ can be chosen  to 
minimize the variance
 $\delta \zeta ^2$ of the random variable  $\zeta $. The optimal choice looks like
\begin{eqnarray}
\beta _j =\frac{1}{\delta \zeta _j^{2}} \Bigl [\sum _{m=1}^M \frac{1}{\delta \zeta _m^{2}}
\Bigr ] ^{-1} 
\label{minbetj}
\end{eqnarray}
and gives the final result for $\delta \zeta ^2$  
\begin{eqnarray}
\delta \zeta  ^2 =\Bigl [\sum _{j=1}^M \frac{1}{\delta \zeta  _j^{2}} \Bigr ] ^{-1} \;.
\label{unczet}
\end{eqnarray}  
Some subtle points concerning properties of $ \zeta$ will be discussed in the 
next section. 

The method discussed above is applicable also for the case when the 
detector efficiency is unstable for 
the time of data taking. The longitudinal double spin asymmetry in 
inclusive or semi-inclusive deep 
inelastic scattering of leptons off nucleons is defined by the relation
\begin{equation}
A_{LL}=\frac{d\sigma _{++}-d\sigma _{+-} }{d\sigma _{++}+d\sigma _{+-} }
\label{all}
\end{equation}  
where $d\sigma _{++}$ ($d\sigma _{+-}$) denotes the differential 
cross section of the studied process
when both helicities of colliding particles are positive 
(have different signs). The typical time 
of data taking for the modern experiments is about few months. 
If the detector efficiency and 
the efficiency of the luminosity monitor changes  considerably during this 
time we cannot use the relations
\begin{equation}
d \sigma _{++}=\frac{n_{++} }{E L_{++}}\;,\; \; \; \;\; 
d \sigma _{++}=\frac{n_{+-} }{E L_{+-}}
\label{siglum}
\end{equation}  
where $E$ denotes as before the detector efficiency, $n_{++}$,
 $n_{+-}$ are observed numbers 
of events  and $L_{++}$,  $L_{+-}$ are integrated luminosities 
when both helicities have the same
sign and opposite signs, respectively. If the direction of the 
target (or beam) polarization changes
for example every run we may ignore all the instabilities 
and define the random variables
$\zeta _j$
\begin{equation}
 \zeta _j=\frac{n_{++}^{2j-1}-n_{+-}^{2j} }{n_{++}^{2j-1}+n_{+-}^{2j}}
\label{runzetj}
\end{equation}  
for two neighbour runs with numbers $(2j-1)$ and $(2j)$. Formula  
$(\ref{runzetj})$ corresponds to relations
$(\ref{defxijzetj1})$, $(\ref{defxijzetj2})$ with $b_j \equiv 1$.
The expected value $<\zeta _j>$ 
for every $\zeta _j$ is equal to the physical asymmetry $A_{LL}$  and hence we 
may apply formulas $(\ref{defzet})$, $(\ref{minbetj})$, 
$(\ref{unczet})$ to minimize the statistical 
uncertainty of the measured asymmetry.  
When writing $(\ref{runzetj})$ we supposed  
that the integrated luminosities are equal 
to each other for the neighbour runs.

\section{Expected value and variance of measured asymmetry} 

Let us consider the random variable $\xi$ defined by $(\ref{randxi})$ and let the 
random variables $p$ and $n$ have  the Poisson probability densities
\begin{equation}
W_p(p)=\frac{x^p}{p!}e^{-x}\;,\;\;\;\;\;W_n(n)=\frac{y^n}{n!}e^{-y}\; 
\label{poiss}
\end{equation}                                      
with the  expected values for $p$ and $n$ being equal to $x$ and $y$, 
respectively. First we consider the case when the background  contribution
is negligible.
Since $x$ ($y$) is proportional to the cross section for the case when the product 
of the initial particle helicities is positive  (negative)  the physical 
asymmetry looks like
\begin{equation}
C=\frac{x-y}{x+y}\;. 
\label{asymc}
\end{equation}
It follows from $(\ref{randxi})$ that $\xi$ becomes  meaningless for $p+n=0$, hence 
we should exclude the term with 
$p+n=0$ in computing its expected value  $<\xi>$. Then the formula for  $<\xi>$ reads
\begin{eqnarray}
\nonumber
<\xi>=\frac{1}{W_+} \sum _{p+n \geq 1}\frac{x^p}{p!}\frac{y^n}{n!} 
\frac{p-n}{p+n}  e^{-x-y}=\\
\nonumber
\frac{x}{W_+} \sum _{p+n \geq 1}\frac{x^{p-1}}{(p-1)!}\frac{y^n}{n!} 
\frac{1}{p+n}  e^{-x-y}- \\
\frac{y}{W_+} \sum _{p+n \geq 1} \frac{x^p}{p!} \frac{y^{n-1}}{(n-1)!}
\frac{1}{p+n}  e^{-x-y}\;. 
\label{mathxi1}
\end{eqnarray}  
Putting  $p+n-1=m$   in the  sums in $(\ref{mathxi1})$  and applying  
the Newton binomial theorem one gets
\begin{eqnarray}
\nonumber
<\xi>=\\
\nonumber
\frac{ e^{-x-y} }{W_+}\sum _{m=0}^{\infty}\frac{1}{m+1} \Bigl[x
 \sum _{n=0}^{m}\frac{x^{m-n}}{(m-n)!}\frac{y^n}{n!}-
y \sum _{p=0}^{m}\frac{x^p}{p!}\frac{y^{m-p}}{(m-p)!} \Bigr]=\\
(x-y)\frac{ e^{-x-y} }{W_+}\sum _{m=0}^{\infty}\frac{(x+y)^m}{(m+1)!}=
\Bigl(\frac{ x-y }{x+y}\Bigr)\frac{1- e^{-x-y} }{W_+}\;.
\label{mathxi2}
\end{eqnarray} 
Since we have excluded in $(\ref{mathxi2})$ the contribution with $p=n=0$ 
we are to divide the sum over $p$ and $n$ by the probability of events with
 $p+n > 0$ which is
\begin{equation}
W_+=1-W_p(0) W_n(0)=1- e^{-x-y}\;.
\label{posprob}
\end{equation}            
Combining $(\ref{mathxi2})$ with $(\ref{posprob})$ we get our final result for 
the first moment of $\xi$
\begin{equation}
<\xi>=\frac{ x-y }{x+y}\;.
\label{mathfin}
\end{equation}            

Comparison of  $(\ref{mathfin})$ with  $(\ref{asymc})$  shows that the 
 expected value of the random 
variable $\xi$ coincides with the physical asymmetry $C$ not only for high 
statistics experiments when
$x \gg 1$, $y \gg 1$ but for any values of $x$ and $y$. The obtained result is rather 
surprising. It allows to
measure the asymmetry in small kinematic regions (bins) and make use of total statistics for 
all the bins where
the  expected values of random variables are equal to each other, to improve  statistical 
accuracy of the obtained 
asymmetry. As has been explained in the Introduction this method allows to avoid 
a study of the detector
efficiency. To realize this program we need the formula for the variance of $\xi$ 
%(defined by $(\ref{randxi})$) 
for every small bin to apply it in formula 
$(\ref{unczet})$.

To calculate the variance of $\xi$ let us consider the second moment $<\xi ^2>$ 
and represent it as the 
sum of four terms $I_1$,   $I_2$,  $I_3$ and  $I_4$ where 
\begin{eqnarray}
\nonumber
<\xi ^2>=\Bigl\langle \frac{(p-n)^2}{(p+n)^2} \Bigr\rangle=
\Bigl\langle \frac{p(p-1)}{(p+n)^2}\Bigr\rangle
+\Bigl\langle \frac{n(n-1)}{(p+n)^2}\Bigr\rangle - \\
%%%%%%\nonumber
2\Bigl\langle \frac{pn}{(p+n)^2}\Bigr\rangle 
+\Bigl\langle \frac{1}{(p+n)}\Bigr\rangle =
I_1+I_2+I_3+I_4\;.
\label{xi2}
\end{eqnarray}  
For calculation of $I_1$ we make use of the relation
\begin{equation}
\frac{1}{(p+n)^q} = \frac{1}{(q-1)!} \int _0^{\infty}e^{-\alpha (p+n)} 
\alpha^{q-1} d \alpha 
\label{gamma}
\end{equation}   
for $q=2$. This gives for $I_1$
\begin{eqnarray}
\nonumber
I_1=\Bigl\langle \frac{p(p-1)}{(p+n)^2}\Bigr\rangle= \\
\nonumber
\frac{1}{W_+} \sum_{p=2}^{\infty}  \sum_{n=0}^{\infty}
\int_0^{\infty}
\frac{x^p}{(p-2)!}\frac{y^n}{n!} e^{-\alpha (p+n)} \alpha d\alpha e^{-x-y}=\\
\nonumber
\frac{x^2 e^{-x-y} }{W_+}\sum_{m=0}^{\infty}  \sum_{n=0}^{m}\int_0^{\infty}
\frac{x^{m-n}}{(m-n)!}\frac{y^n}{n!} e^{-\alpha (m+2)} \alpha d\alpha =\\
\frac{x^2}{W_+} e^{-x-y} \int_0^{\infty} \alpha e^{-2 \alpha } 
\exp\{(x+y) e^{- \alpha } \}  d\alpha \;
\label{i1}
\end{eqnarray}  
where $W_+$ has been defined in $(\ref{posprob})$ and $m=p+n-2$.
After integrating by parts integral $(\ref{i1})$ can be represented as
\begin{equation}
I_1= \frac{x^2}{W_+ (x+y)}[1-e^{-x-y}-\phi(x+y)]
\label{i1phi}
\end{equation}   
where we denote by $\phi(z)$ the function
\begin{equation}
\phi(z)=e^{-z}\int_0^{z} \frac{e^t-1}{t} dt = e^{-z} \sum_{m=1}^{\infty} \frac{z^m}
{m \cdot m!}\;.
\label{phiser}
\end{equation}   
The integrals $I_2$ and $I_3$ can be calculated in an analogous way and we get
\begin{equation}
I_1+I_1+I_3= \frac{(x-y)^2}{(x+y)^2} 
\Bigl [1-\frac{\phi(x+y)}{1-e^{-x-y}}\Bigr ]\;.
\label{i1i2i3}
\end{equation}   
We have taken into account  $(\ref{posprob})$ in $(\ref{i1i2i3})$.
The calculation of $I_4$ is trivial. Indeed, remembering $(\ref{posprob})$ and 
introducing $m=p+n$ one has
\begin{eqnarray}
\nonumber
I_4=\Bigl\langle \frac{1}{p+n}\Bigr\rangle = \\
\nonumber
\frac{1}{W_+}\sum_{p+n \geq 1}^{\infty}
\frac{x^p}{p!}\frac{y^n}{n!}\frac{ e^{-x-y}}{p+n}=
\frac{ e^{-x-y}}{W_+}\sum_{m=1 }^{\infty}\frac{1}{m}\sum_{n=0 }^{m} 
\frac{x^{m-n}}{(m-n)!}\frac{y^n}{n!}= \\
\frac{ e^{-x-y}}{W_+}\sum_{m=1 }^{\infty} \frac{(x+y)^m}{m \cdot m!}=
\frac{ \phi (x+y)} {1- e^{-x-y}}
\label{i4} 
\end{eqnarray}   
where $\phi(z)$ is defined by $(\ref{phiser})$. Putting  $(\ref{i1i2i3})$ and
$(\ref{i4})$ into $(\ref{xi2})$ and remembering $(\ref{mathfin})$
we get easily the final formula for the variance of $\xi$ 
\begin{equation}
\delta \xi ^2=<\xi ^2> -<\xi > ^2 =
\frac{\phi (x+y)}{1- e^{-x-y}} \Bigl [1-\frac{(x-y)^2}{(x+y)^2} \Bigr ] \;.
\label{uncerfin}
\end{equation}   
For numerical calculations with  formula $(\ref{uncerfin})$,
it is convenient to use the integral representations for $\phi(z)$ 
\begin{eqnarray}
\phi(z)=-\int_0^{z} \ln(1-t/z)e^{-t} dt\;, 
\label{intrephi1} 
\end{eqnarray} 
\begin{eqnarray}
\phi(z)=1-e^{-z}+z^2\int_0^{1}(1-t) \ln (1-t) e^{-zt} dt  \;.
\label{intrephi2} 
\end{eqnarray} 

Now we are going to discuss some properties of the variance 
$(\ref{uncerfin})$. For a high statistics case when $x+y \gg 1$ we may use the 
asymptotic power series for $\phi(z)$ valid if $z \rightarrow \infty$
\begin{equation}
\phi(z)=\sum_{n=0 }^{\infty}\frac{n!}{z^{n+1}} \;.
\label{asymphi}
\end{equation}   
Formula $(\ref{asymphi})$ can be easily derived at $z \gg 1$ from relation 
$(\ref{intrephi1})$ through  decomposition of $\ln(1-t/z)$ into a power 
series and 
integration over $t$ with the limits $0$ and $\infty$. Substituting $(\ref{asymphi})$ into 
 $(\ref{uncerfin})$ we get the asymptotic relation valid at $x+y \gg 1$
\begin{equation}
\delta \xi ^2 =\frac{1-C^2}{x+y} + O((x+y)^{-2})
\label{asymsigma}
\end{equation}   
where $C$ is the physical asymmetry defined by $(\ref{asymc})$ and
\linebreak 
$O((x+y)^{-2})$ denotes terms of the order $(x+y)^{-2}$ and smaller at $x+y 
\rightarrow \infty$. Since $x$ and $y$ are 
the expected values of observed numbers of events $p$ and $n$, respectively,
formula $(\ref{asymsigma})$ says that
$\delta \xi$ decreases as reciprocal of a square root of the total event number. The solid line in 
Fig.~1 shows the dependence of $\phi(z)$ on $z$, and the dash-dotted curve represents
a dependence on $z=x+y$ of the ratio of the variance calculated with  
$(\ref{uncerfin})$ to its asymptotic expression given by $(\ref{asymsigma})$. We see 
that the ratio deviates significantly from the unity at $z \leq 15$ especially at
$z \sim 5$. This means that widely used formula $(\ref{asymsigma})$ for the variance 
of the measured asymmetry is not valid for the low statistics. 
We would like
to stress that though $\xi$ is the ratio of the random variables $p-n$ and $p+n$
(see$ (\ref{randxi})$) it has a finite statistical uncertainty at any $x$ and $y$ which
follows immediately 
from $(\ref{uncerfin})$,  $(\ref{intrephi2})$ and
 $(\ref{asymsigma})$. This result is in a contrast with the well known property of
the Cauchy random variable having the infinite second moment. The reason for this is the
used convention of removing the contribution of events with $p+n=0$. Another
 difference of principle  between the Cauchy distribution and the probability
distribution of $\xi$  
 is as follows.
Since for $\xi$ defined by $(\ref{randxi})$ the numerator $p-n$ and the denominator
$p+n$ are correlated we have for any integer numbers $p$, $n$ and $k$ a relation
\begin{eqnarray}
-1 \leq \Bigl (\frac {p-n}{p+n} \Bigr )^k  \leq 1
\label{uneq11}   
\end{eqnarray}
which means that all moments of $\xi$ are finite if $p+n >0$. The Cauchy random variable
 is a ratio of two uncorrelated random variables.
We see also from $(\ref{uncerfin})$ and $(\ref{asymc})$  that $\delta \xi ^2 =0$ if 
\linebreak
$|C|=1$. This property of
$\delta \xi ^2 $ is due to the  fact that for $x=0$ the probability to observe $p>0$
vanishes  according to $(\ref{poiss})$, hence $\xi=-1$ for any
$n$, which means $\delta \xi^2  =0$.  It follows also from $(\ref{poiss})$  that $n \equiv 0$ at
$y=0$,
hence $\xi=1$ for any $p$.
The property under
discussion can be proved from the obvious inequality
\newpage
%%%%\linebreak
\begin{center}
\mbox{\epsfig{file=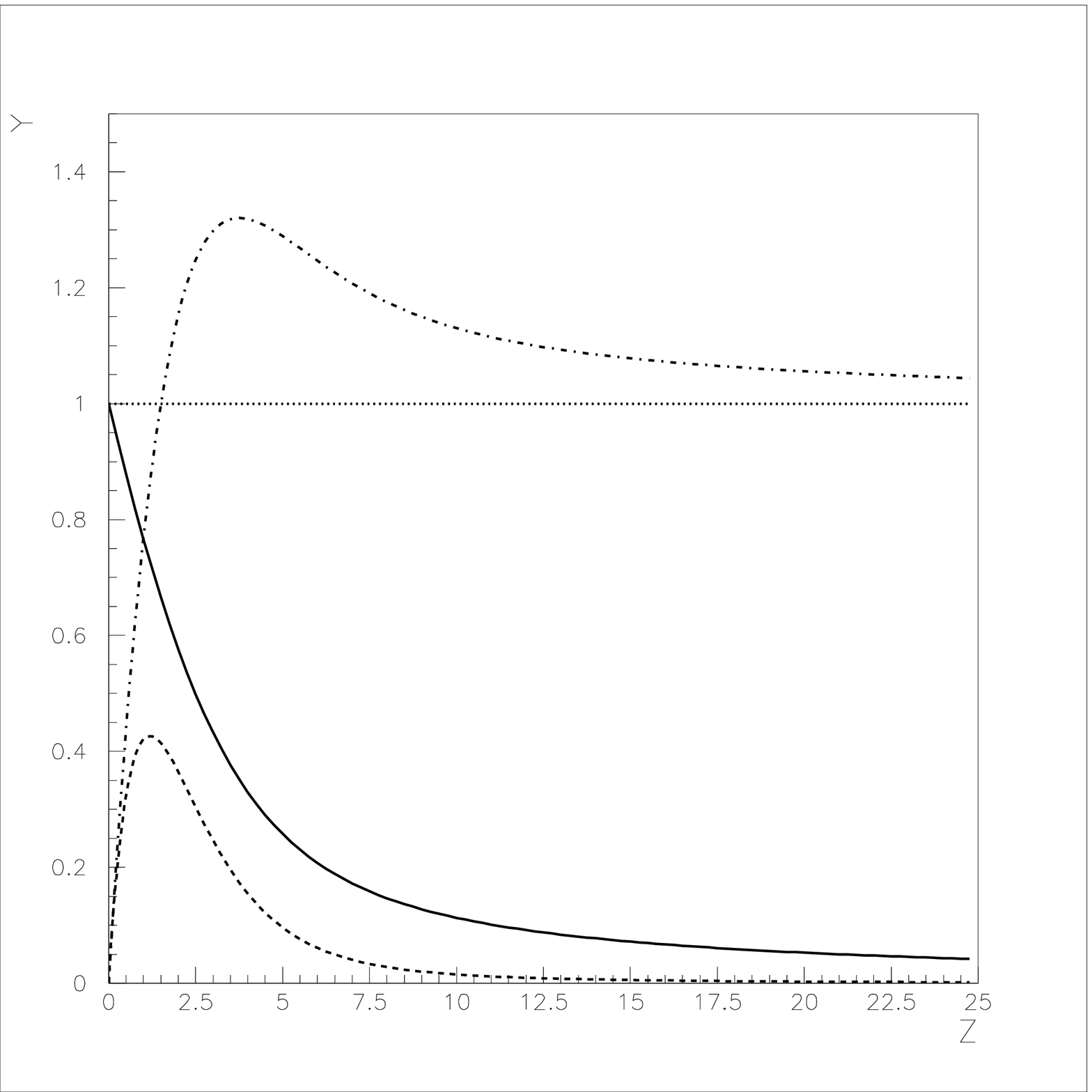,height=18cm,width=15cm}}
%angle=270}}   XS
\noindent
\small
\begin{minipage}{14.cm}
\vspace{7mm}  
{\sf Fig.~1:}
Functions $\phi(z)$, and $\chi(z)$.
Solid and dashed curves show functions $\phi(z)$,  $\chi(z)$ defined by $(\ref{intrephi2})$,
$(\ref{defchi})$, respectively.
Dash-dotted curve represent the function $z*\phi(z)$.
\end{minipage}
\end{center}
\vspace{0.3cm}
\normalsize  
\begin{eqnarray}
\nonumber
1 \geq <\xi ^2> =<[(\xi- <\xi>)+<\xi>]^2>=\delta \xi ^2+<\xi>^2 
\end{eqnarray}   
which means that
\begin{equation}
\delta \xi ^2 \leq 1-<\xi>^2 =1-C^2\;.
\label{uneq2}
\end{equation}   

Let us consider the limit of expression $(\ref{uncerfin})$ at $x+y \rightarrow 0$ which 
will be used below. Applying $(\ref{intrephi2})$  at low $x+y$ we have
\begin{equation}
\phi (x+y)= x+y+ O((x+y)^2)
\label{zerphi}
\end{equation}   
and substituting $(\ref{zerphi})$ into $(\ref{uncerfin})$ we get the relation valid at 
\linebreak
$x+y \rightarrow 0$
\begin{equation}
\delta \xi ^2 =1-\Bigl (\frac{x-y}{x+y} \Bigr )^2= 1-<\xi>^2 
\label{lowxeq}
\end{equation}   
which realizes the upper limit in inequality $(\ref{uneq2})$.

As has been discussed in the Introduction, to increase a statistical accuracy of 
the measured physical quantity (the polarization, the spin-spin asymmetry etc.) we are 
to utilize total number of experimental events. For this reason we consider all those bins
in which the first moments of $\zeta _j$ defined by  $(\ref{defxijzetj1})$, $(\ref{defxijzetj2})$
 are equal to each other. 
But in the practical application of formulas   $(\ref{defzet})$,  $(\ref{minbetj})$,
 $(\ref{unczet})$ we should remember that $\xi_j$ and $\zeta_j$ are not defined for those bins where 
$p_j+n_j=0$. Hence we may  use  $M$ bins among $N$ bins only $(M \leq N)$. 
It is obvious that $M$ cannot
be larger than the observed number of the experimental events $N^{exp}$. It is easy 
to get the upper limit for the variance of $\zeta$ defined by 
$(\ref{defzet})$ and $(\ref{minbetj})$. Let us consider
  very small bins so that $N \gg N^{exp}$. For this case the expected values of the observed 
numbers of events $p_j$ and $n_j$ are much less than unity $(x_j \ll 1,\;y_j \ll 1)$ 
and $M$ coincides 
practically with  $N^{exp}$. We may apply $(\ref{lowxeq})$ for the variance of $\xi_j$. 
Remembering the relation between $\xi _j$ and $\zeta _j$ given by $(\ref{defxijzetj2})$ we get 
the following formula for the variance of $\zeta _j$ valid if $x_j \ll 1,\;y_j \ll 1$
\begin{equation}
\delta \zeta^2_j =\frac{1}{b^2_j}(1-<\xi _j>^2) =\frac{1}{b^2_j}(1- b^2_j <\zeta _j>^2)\;.             
\label{varzetaj}
\end{equation}   
Taking into account that all  $<\zeta _j>$  are equal to   $<\zeta >$ 
we substitute $(\ref{varzetaj})$ into $(\ref{unczet})$ and come to the relation 
\begin{equation}
\delta \zeta ^2 = \Bigl [\sum_{j=1}^M\frac{b^2_j}{1- b^2_j <\zeta >^2} \Bigr ]^{-1}\;.             
\label{inqvarz}
\end{equation}   
Since for physically interesting cases $0 \leq b^2_j <\zeta >^2 < 1$ we get from $(\ref{inqvarz})$ 
the inequality of interest
\begin{equation}
\delta \zeta ^2 \leq 
\frac{1}{M <b^2>_M}             
\label{inqvint}
\end{equation}   
where $<b^2>_M$ is the arithmetic mean of $b^2_j$ over $M$ bins
\begin{equation}
<b^2>_M= \frac{1}{M} \sum_{j=1}^M b^2_j\;.             
\label{defarmean}
\end{equation}   
Since $M \approx N^{exp}$ formula $(\ref{inqvint})$ shows that even for very small bins when 
observed event numbers $p_j$, $n_j$ are about $1$ we can get high 
statistical accuracy $(\delta \zeta \sim 1/\sqrt{N^{exp}})$ if 
$N^{exp} \gg 1$.

The procedure of extraction of the physical quantity $<\zeta>$ and 
its variance $\delta \zeta ^2$ from experimental data can be as follows.
First, we calculate the arithmetic mean $\bar{\zeta}$ with the aid of the formula
\begin{equation}
\bar{\zeta} =  \frac{1}{M} \sum_{j=1}^M \zeta_j            
\label{armeanzeta}
\end{equation}       
where $\zeta_j$ has been defined by $(\ref{defxijzetj1})$ and $(\ref{defxijzetj2})$. The
arithmetic mean $\bar{\zeta}$ is an estimate of the physical quantity $<\zeta>$. After that we may
estimate parameters $x_j$ and $y_j$ putting 
\begin{equation}
x_j= p_j,\; \; \;   y_j= n_j         
\label{estxjyj}
\end{equation}       
and apply  $(\ref{estxjyj})$ to calculate  $\delta \bar{ \zeta }^2_j$ with the aid of 
relation $(\ref{uncerfin})$
\begin{equation}
\delta \bar{ \zeta}^2_j=  \frac{\phi(p_j+n_j)}{b^2_j( 1-e^{-p_j-n_j})}  [1-b^2_j
\bar{\zeta}^2]    
\label{estbarsigj}
\end{equation}       
which is an estimate of the variance of $\zeta _j$. Finally, we put $\delta \bar{ \zeta}^2_j$ 
into $(\ref{unczet})$ instead of $\delta \zeta ^2_j$ to get an estimate of the variance of
$\zeta $ where
\begin{equation}
\delta \bar{ \zeta} ^2 =\Bigl [\sum _{j=1}^M 1/ \delta \bar{ \zeta} _j^{2} \Bigr ] ^{-1} \;.        
\label{finestsig}
\end{equation}

Since $p$ and $n$ are integer numbers the random variable $\xi$ is defined for rational numbers in
accordance with $(\ref{randxi})$. The probability $W_{\xi}$ to observe $\xi=(p-n)/(p+n)$ is not a
smooth function of $\xi$. We may formally define the quantity $\Omega$ by the relation
\begin{equation}
\Omega(\xi,h)=\frac{1}{h}W_{\xi}(\xi-\frac{h}{2} < \frac{p-n}{p+n} < \xi+\frac{h}{2})
\label{densomeg}
\end{equation}
which would become the probability density in the limit $h \rightarrow 0$ if $W_{\xi}$ were a
smooth function . Figure~2 shows the
dependence of $\Omega$ on $\xi$ at $x=6$, $y=2$ for numbers of bins $N$ equal to 30 (top) and
$N=150$ (bottom) where $h$ and $N$ in $(\ref{densomeg})$ are related with the formula 
$h=2/N$. It is obvious that
there is no  smooth function $\Omega$ given by $(\ref{densomeg})$ at $h \rightarrow 0$. Figure~3
shows that $\Omega$ corresponds to more smooth histograms than those presented in 
Fig.~2 if we
increase $x$ and $y$ keeping the asymmetry $C$ (see $(\ref{asymc})$) equal to the same value as in
Fig.~2. Figure~3 corresponds to higher statistics in comparison with Fig.~2 since $x=60$, $y=20$.
It is easy to see from Fig.~3 that  for high statistics  $\Omega$
resembles the Gaussian distribution.

To find the confidence levels which correspond to
the regions
\begin{equation}
C-k \delta \xi  <  \xi < C+k \delta \xi
\label{conflev}
\end{equation}
($k=1,\;2,\;3$) usually used in the statistical analysis of the experimental data we are to
calculate the cumulative distribution using the formula
\begin{equation}
F(z)=W_{\xi}(\xi <z)=\sum_{\xi <z} \frac{x^p}{p!}\frac{y^n}{n!} e^{-x-y}/W_+\;.
\label{cumdistr}
\end{equation}
with $W_+$ given by  $(\ref{posprob})$.
The sum in $(\ref{cumdistr})$ runs over all positive integer $p$ and $n$ obeying the unequality $z
> \xi= (p-n)/(p+n)$. The values of $<\xi>$, $\delta \xi ^2$ are calculated with the aid of
$(\ref{mathfin})$,  $(\ref{uncerfin})$ and to compute $\omega (k \delta \xi )$
($k=1,\;2,\;...\;$)

\begin{equation}
\omega(k \delta \xi  )=F(<\xi>+k \delta \xi )-F(<\xi>-k \delta \xi )
\label{defconflev}
\end{equation}
relation $(\ref{cumdistr})$ has been used. The results of the calculations are presented in
Table~1.
\linebreak
\newpage
%%%\vspace{0.5cm}
\begin{center}
\begin{tabular}{c}
\mbox{\epsfig{file=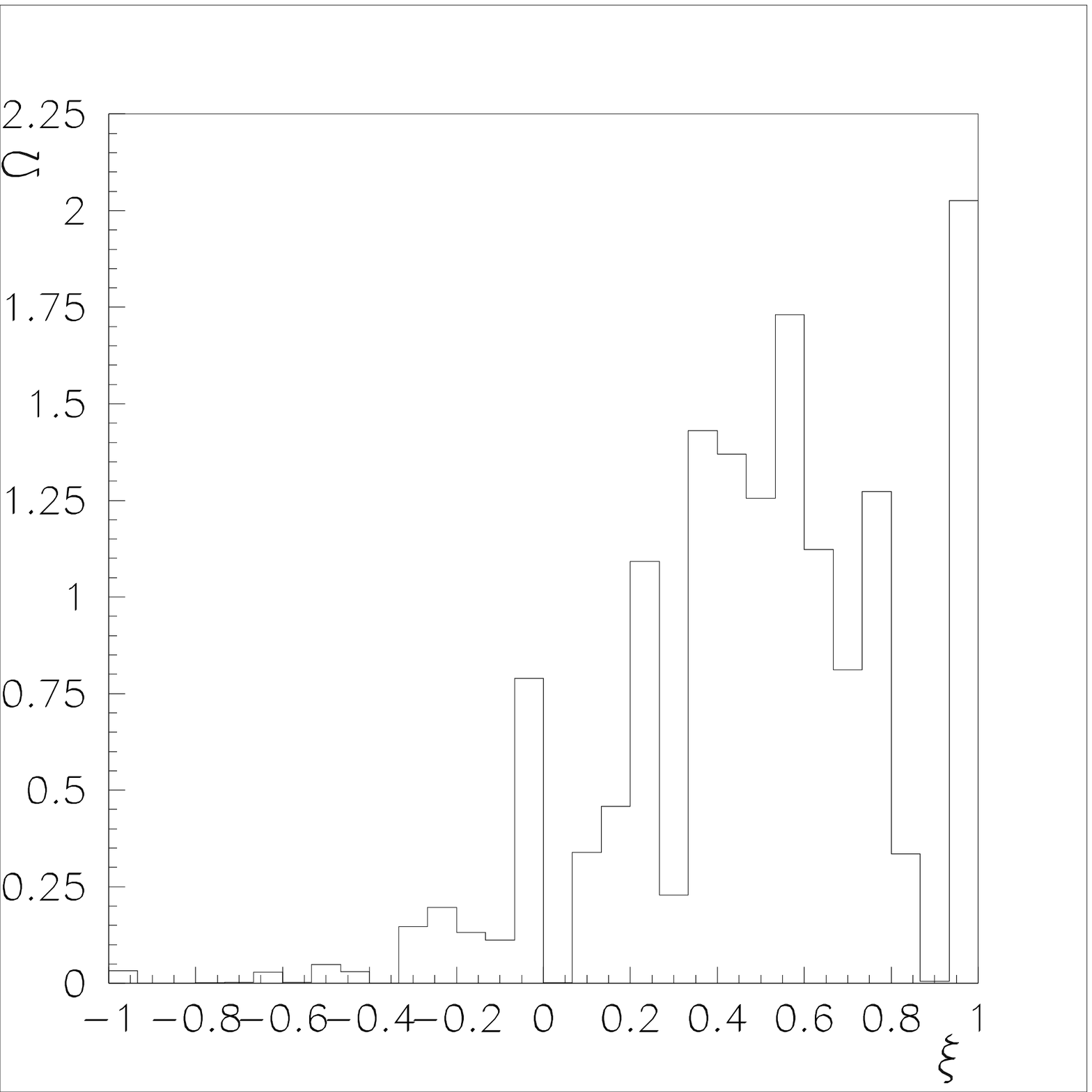,height=10.5cm,width=10.0cm}}
\vspace{5mm}
\noindent
\small
\\

\mbox{\epsfig{file=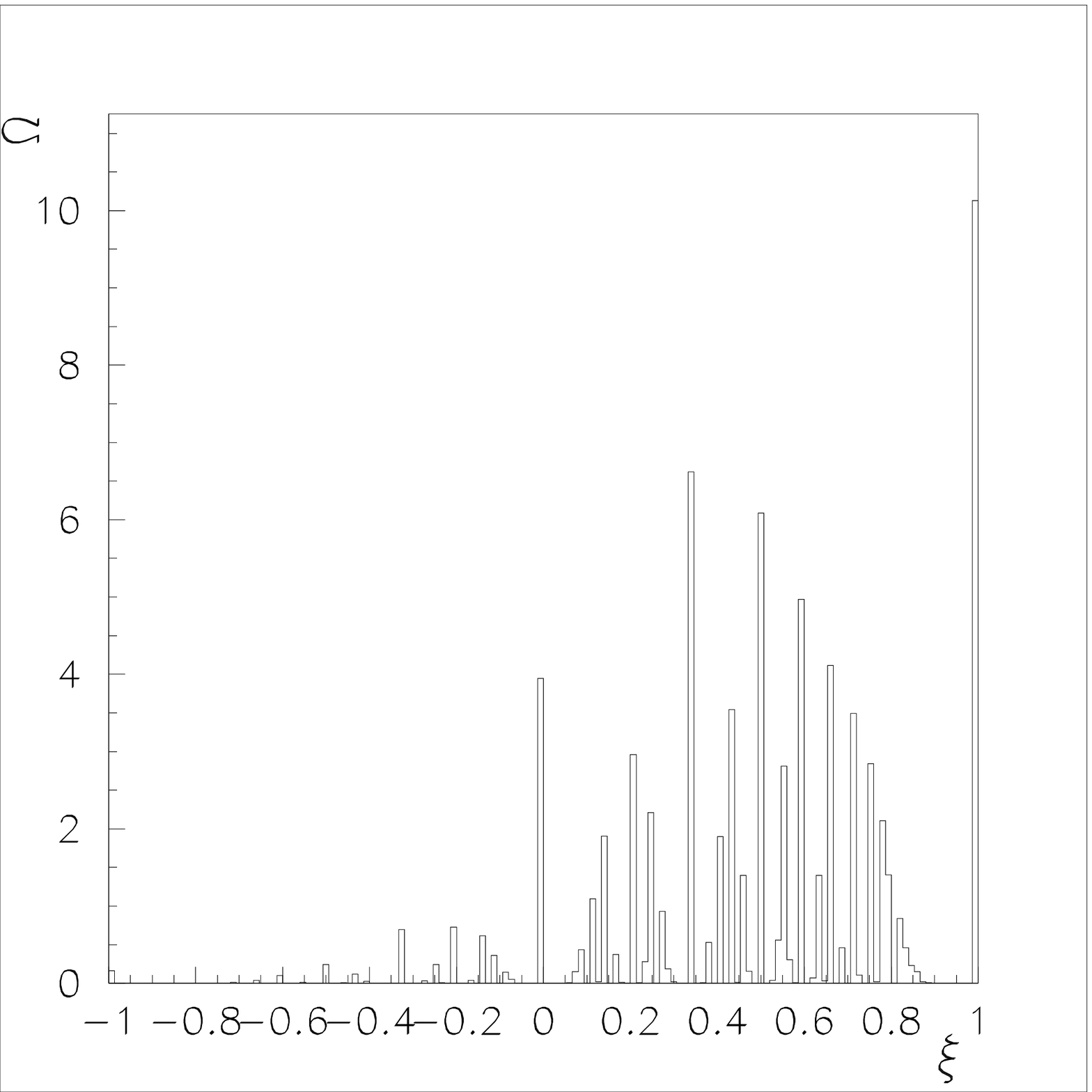,height=10.5cm,width=10.0cm}}
\vspace{5mm}  
\noindent
\small
\\
\begin{minipage}{14cm}
{\sf Figure~2:}
Probability density $\Omega$ for random variable $\xi$. Histograms are computed for $x=6$ and $y=2$.
Bin number: 
%%%\linebreak
top - $N=30$, bottom - $N=150$.
\end{minipage}
 \normalsize  
\end{tabular}
\end{center}
%%\newpage
%%%\vspace{0.5cm}
\begin{center}
\begin{tabular}{c}
\mbox{\epsfig{file=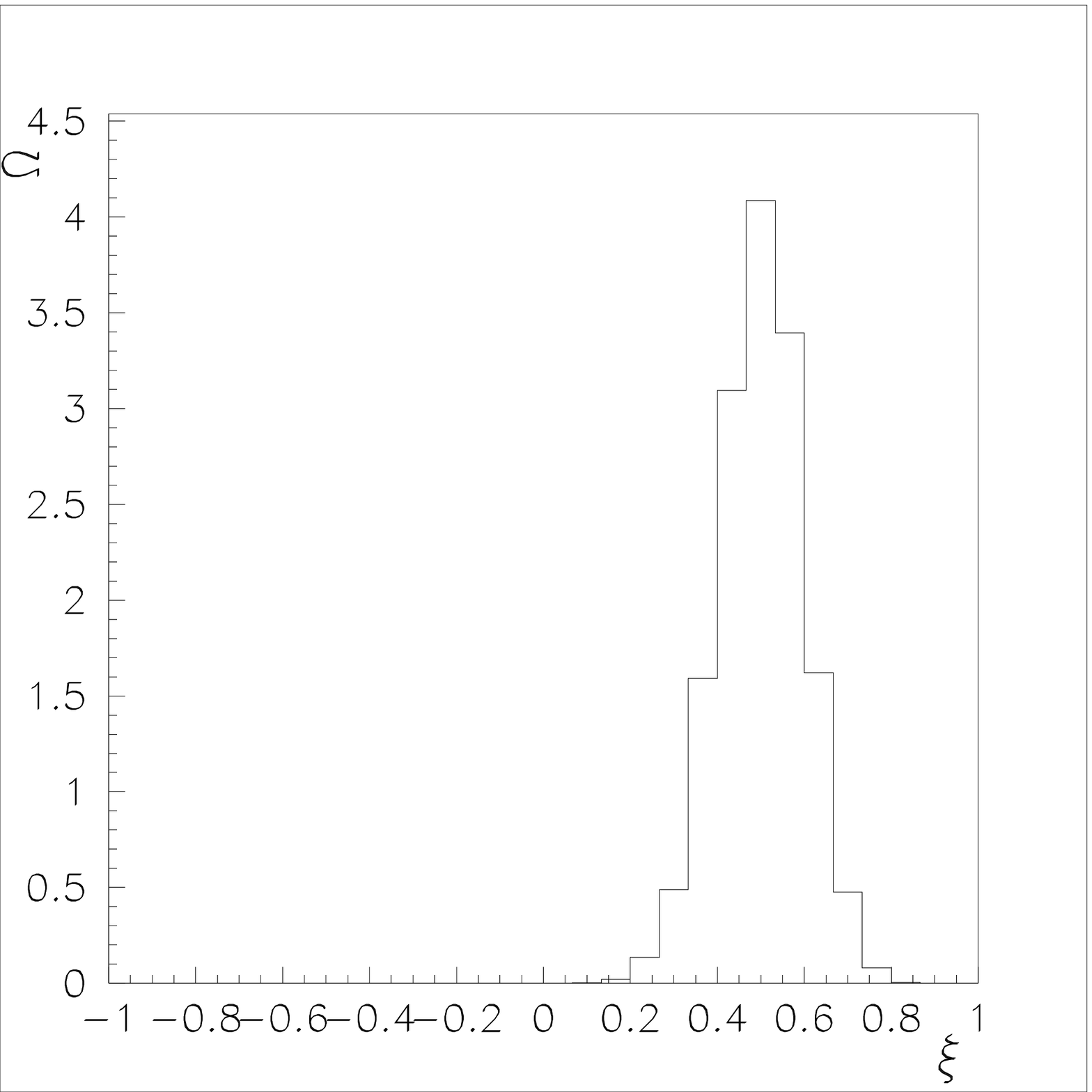,height=10.5cm,width=10.0cm}}
\vspace{5mm}
\noindent
\small
\\
\mbox{\epsfig{file=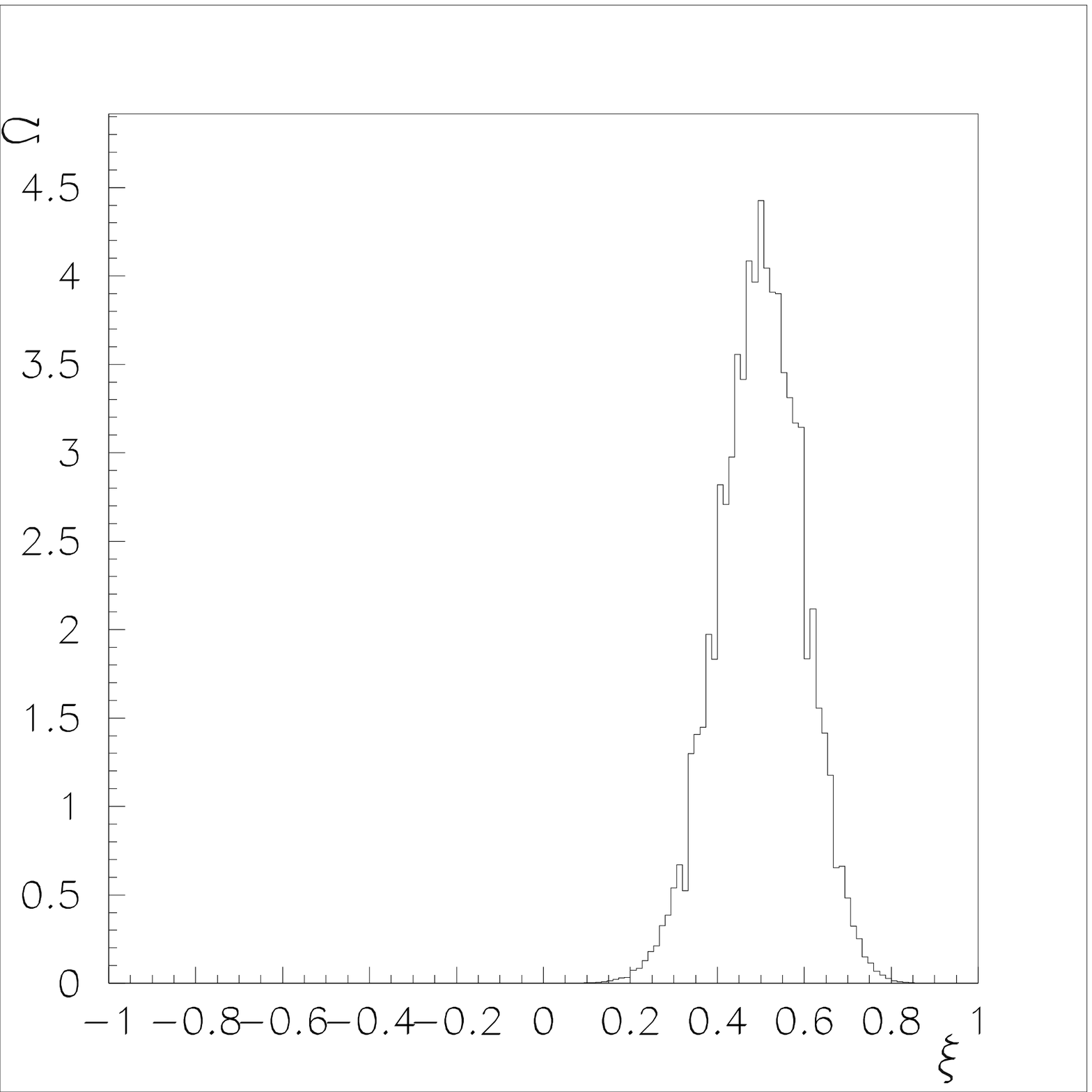,height=10.5cm,width=10.0cm}}
\vspace{5mm}
\noindent
\small
\\
\begin{minipage}{14cm}
{\sf Figure~3:}
Probability density $\Omega$ for random variable $\xi$. Histograms are computed  for $x=60$ and $y=20$.
Bin number: 
%%%\linebreak
top - $N=30$, bottom - $N=150$
\end{minipage}
\normalsize
\end{tabular}  
\end{center}
The first line in Table~1 corresponds to the
Gaussian distribution. We see that even at rather low
statistics ($x+y \sim 100$) the numbers $\omega (k \delta \xi )$ agree reasonably with the 
confidence levels for the  normal density function. We see from the
comparison of every line with the first line that the
higher statistics the better is the agreement between the confidence levels for the random variable
$\xi$ with those for the Gaussian distribution. It is easy to see from Table~1 that
the agreement between the confidence levels becomes worse with an increase of $|<\xi>|$
(other things being equal) when the
expected value goes close to its limits equal to $\pm 1$.
 \vspace{0.5cm} 
\begin{center}   
\begin{tabular}{ccccccc}
x&y&$<\xi>$& $\delta \xi $& $\omega(\delta \xi )$&$\omega(2\delta \xi )$&
$\omega(3\delta \xi )$\\
&&&&0.68268&0.94601&0.99730\\
3&1&0.5&0.49712&0.43339&0.95559&0.96794\\
12&4&0.5&0.2242&0.68386&0.95436&0.99615\\
120&40&0.5&0.068682&0.68344&0.95427&0.99720\\
7&1&0.75&0.2535&0.87089&0.96202&0.99453\\
28&4&0.75&0.11886&0.68360&0.95069&0.99546\\
140&20&0.75&0.052456&0.68165&0.95492&0.99693\\
19&1&0.9&0.10016&0.85166&0.95990&0.99207\\
76&4&0.9&0.049045&0.68116&0.94877&0.99527\\
380&20&0.9&0.021822&0.68228&0.95547&0.99694\\
\end{tabular}
\end{center}
\vspace{0.5cm}

%%%%%%%%%%%%%%%%%%%%%%%%%%%%%%%%%%%%%%%%%%%%%%%%%%%%%%%%%%%%%%%%%%%%%%%%%%%%%%%%%%%%%%%%%%%%%
\section{Taking into account background 
\newline
 events}

Usually measured numbers of events consist of events of the process under 
investigation and background events.
Let the background events are described with random variables $m$, 
$k$ both having the Poisson distributions
\begin{equation}
W_m(m)=\frac{z^m}{m!}e^{-z}\;,\;\;\;\;\;W_k(k)=\frac{t^k}{k!}e^{-t}\; 
\label{backpoiss}
\end{equation}                                      
where $m$ ($k$) describes the background contribution to the observed number $p$ ($n$). 
Since the expected numbers
of the events for the process under investigation are equal to $x-z$ and $y-t$, 
then the formula for 
the physical asymmetry reads
\begin{equation}
C=\frac{x-z-y+t}{x+y-z-t}\;.        
\label{asymback}
\end{equation}        
We suppose as before that the region of the kinematic variables is small enough 
therefore the detector efficiency can be considered as a constant and hence the ratio 
in $(\ref{asymback})$ does not depend on the detector efficiency.

We can obtain the asymmetry if
\begin{equation}
p+n-m-k \neq 0        
\label{reg01}
\end{equation}        
but we start our discussion considering the measured asymmetry in case when a total number of 
events $p+n$ is larger than a total number of background events $m+k$
\begin{equation}
p+n-m-k > 0\;.        
\label{reg02}
\end{equation}
The total probability to observe events in region $(\ref{reg02})$ will 
be denoted as $W_+$ and is given by
the relations
\begin{equation}
W_+=1-W_-\;,
\label{winreg11}
\end{equation}
\begin{eqnarray}
\nonumber
W_-=\sum _{p+n \leq m+k} \frac{x^p\; y^n\; z^m\; t^k}{p!\; n!\; m!\; k!} e^{-x-y-z-t}=\\
\sum _{l=0}^{\infty} \frac{(z+t)^l}{l!} \sum _{q=0}^{l} \frac{(x+y)^q}{q!}\varepsilon  =
\sum _{r=0}^{\infty}  \sum _{q=0}^{\infty} \frac{(z+t)^{q+r} (x+y)^q}{(q+r)!\; q!} 
\varepsilon \;.     
\label{winreg02}
\end{eqnarray}
For transformation of the sums in $(\ref{winreg02})$ we put $q=p+n$, $l=m+k$ 
and after that we define $r=l-q$. 
We have applied in $(\ref{winreg02})$ the short notation
\begin{equation}
\varepsilon=e^{-x-y-z-t}\;.        
\label{epsilon}
\end{equation}
Making use of the well known formula \cite{Whitt}
\begin{equation}
\sum _{r=0}^{\infty} \frac{\xi^r}{r!(\nu+r)!}=\xi^{-\nu /2} I_{\nu}(2\sqrt{\xi})   
\label{bessel01}
\end{equation}
for the Bessel functions $I_m(z)=(-i)^m J_m(iz)$ depending on the imaginary argument 
we get from $(\ref{winreg02})$
\begin{equation}
W_-=\sum _{r=0}^{\infty}\Bigl (  \frac{z+t}{x+y} \Bigr )^{r/2}
 I_r\Bigl (2\sqrt{(x+y)(z+t)}\Bigr )e^{-x-y-z-t}  \;.        
\label{wminus}
\end{equation}
Remembering another well known formula for the Bessel functions \cite{Whitt}
\begin{equation}
J_n(z)=  \frac{1}{2 \pi i} \oint u^{-n} \exp \{\frac{z}{2}(u+\frac{1}{u})\} \frac{d u}{u}         
\label{intbess}
\end{equation}
and putting
\begin{equation}
u=a\sqrt{  \frac{x+y}{z+t}}  \exp \{i( \phi-\frac{\pi}{2}) \}        
\label{uviaxy}
\end{equation}
with arbitrary  $a$ obeying the inequality 
\begin{equation}
a >1        
\label{agreat1}
\end{equation}
we get the useful expression 
\begin{eqnarray}
\nonumber
I_n(2\sqrt{(x+y)(z+t)})=\\
  \frac{a^{-n}}{2 \pi } \Bigl (\frac{z+t}{x+y}\Bigr )^{n/2} 
\int_0^{2 \pi} d \phi e^{-in \phi}
E(x+y,z+t,\phi)         
\label{bessel02}
\end{eqnarray}
with
 \begin{eqnarray}
E(x+y,z+t,\phi)= \exp \{ a(x+y)e^{i \phi}+ \frac{1}{a}(z+t)e^{-i \phi}\}\;.
\label{EXYZT}
\end{eqnarray}
Summing over $r$ in $(\ref{wminus})$ with the aid of $(\ref{bessel02})$ 
and remembering $(\ref{winreg11})$
we obtain the final result for $W_+$
 \begin{eqnarray}
\nonumber
W_+=1-
  \frac{\varepsilon }{2 \pi } \int_0^{2 \pi} d \phi 
E(x+y,z+t,\phi)\\
\Bigl [1-\frac{z+t}{a(x+y)}e^{-i \phi} \Bigr ] ^{-1} d \phi \;.         
\label{finwplus}
\end{eqnarray}

Let us define the random variable $\eta$ by the relation
\begin{equation}
\eta=\frac{(p-m)-(n-k)}{(p-m)+(n-k)}        
\label{difeta}
\end{equation}
and consider its expected value in region  $(\ref{reg02})$
 \begin{eqnarray}
\nonumber
<\eta>W_+=\varepsilon  \sum_{p+n>m+k} \Bigl (\frac{x^p\; y^n\; z^m\; t^k}{p!\; n!\; m!\; k!}\Bigr )
\Bigl ( \frac{p-n-m+k}{p+n-m-k}\Bigr )  = \\
U_p-U_n-U_m+U_k \;,         
\label{upk}
\end{eqnarray}
where for $s=p,\;n,\;m,\;$ or $k$
\begin{equation}
U_s=\varepsilon  \sum_{p+n>m+k} \Bigl (\frac{x^p\; y^n\; z^m\; t^k}{p!\; n!\; m!\; k!}\Bigr )
\Bigl ( \frac{s}{p+n-m-k}\Bigr ) \;.       
\label{us}
\end{equation} 
%%The quantity $\varepsilon$ has been defined in $(\ref{epsilon})$.
Making use of the obvious relation
\begin{equation} 
\frac{1}{(p+n-m-k)^j} = \frac{1}{(j-1)!} \int _0^{\infty}e^{-\alpha (p+n-m-k)} 
\alpha^{j-1} d \alpha 
\label{jback}
\end{equation}   
at $j=1$ (valid in region $(\ref{reg02})$) we get for $U_p$
 \begin{eqnarray}
\nonumber
U_p= \varepsilon  \int_0^{\infty} d \alpha
 \sum_{p+n>m+k} \frac{x^p\; y^n\;   z^m\;  t^k}{(p-1)!\; n!\; m!\; k!} e^{-\alpha(p+n-m-k)}  =\\
\nonumber
x\varepsilon  \int_0^{\infty}  d \alpha  e^{-\alpha}
 \sum_{s+n \geq m+k} \frac{x^s\; y^n\; z^m\; t^k}{s!\; n!\; m!\; k!} e^{-\alpha(s+n-m-k)}  = \\
\nonumber
x \varepsilon \int_0^{\infty}  d \alpha  e^{-\alpha}
 \sum_{q=0}^{\infty} \frac{(x+y)^q }{q!}\sum_{l=0}^{q} \frac{(z+t)^l }{l!}
e^{-\alpha(q-l)}  =\\
%\nonumber
x \varepsilon \int_0^{\infty}  d \alpha  e^{-\alpha}
 \sum_{r=0}^{\infty}e^{-\alpha r} \sum_{l=0}^{\infty}
\frac{(x+y)^{l+r}(z+t)^l }{(l+r)!\; l!}  \;.         
\label{up01}
\end{eqnarray}
In  $(\ref{up01})$ we put $s=p-1$ and after that  introduce $q=s+n$, $l=m+k$. 
We change the order of 
summation in $(\ref{up01})$ making use of $r=q-l$. We get the chain of the equalities
substituting  $(\ref{bessel01})$ and  then $(\ref{bessel02})$ into $(\ref{up01})$
 \begin{eqnarray}
\nonumber
U_p=x \varepsilon \int_0^{\infty}  d \alpha  e^{-\alpha}
 \sum_{r=0}^{\infty}e^{-\alpha r} \Bigl (\frac{x+y}{z+t}  \Bigr )^{\frac{r}{2}}
I_r(2\sqrt{(x+y)(z+t)})=\\
\nonumber
x \varepsilon \int_0^{\infty}  d \alpha  e^{-\alpha} 
\int_0^{2 \pi} \frac{d \phi}{2 \pi} \sum_{r=0}^{\infty} e^{- r \alpha}  e^{-i r \phi} a^{- r }
E(x+y,z+t,\phi)=\\
%%%%\nonumber
\frac{x \varepsilon }{2 \pi}\int_0^{2 \pi}d \phi \int_0^{\infty}  d \alpha  
\Bigl ( \frac{e^{-\alpha} }{1-e^{-\alpha}e^{-i \phi}/a} \Bigr )
E(x+y,z+t,\phi)
\label{up02}
\end{eqnarray}
where $E(x+y,z+t,\phi)$ has been defined by $(\ref{EXYZT})$. 
The sum over $r$ in $(\ref{up02})$ is convergent due to $(\ref{agreat1})$.  
Integrating over $\alpha$  in $(\ref{up02})$ we get the relation
 \begin{eqnarray}
U_p=\frac{a x \varepsilon }{2 \pi} \int_0^{2 \pi} d \phi e^{i \phi} \ln \Bigl 
(\frac{1 }{1- e^{-i \phi}/a} \Bigr )
E(x+y,z+t,\phi)\;.
\label{up03}
\end{eqnarray}
The formula for $U_n$ can be obtained from  $(\ref{up03})$ by the change $x \leftrightarrow y$.
%%%$y \rightarrow x$. 
The analogous calculation gives
 \begin{eqnarray}
U_k=\frac{t \varepsilon }{2 \pi a} \int_0^{2 \pi} d \phi e^{-i \phi} \ln \Bigl 
(\frac{1 }{1-e^{-i \phi}/a} \Bigr )
E(x+y,z+t,\phi)\;.
\label{uk01}
\end{eqnarray}
The relation for $U_m$ follows from $(\ref{uk01})$ if we make use of the transformations
$t \leftrightarrow z$.
%%%%%$z \rightarrow t$.
Putting 
%$(\ref{finwplus})$,  
$(\ref{up03})$, $(\ref{uk01})$ and relations for $U_n$, $U_m$ into 
$(\ref{upk})$ we get the final result for the expected value of 
the random variable $\eta$ valid if $(\ref{reg02})$ is fulfilled
 \begin{eqnarray}
\nonumber
<\eta>= 
\frac{ \varepsilon }{2 \pi W_+} \int_0^{2 \pi} d \phi
\Bigl [a(x-y)e^{i \phi}+\frac{t-z}{a}e^{-i \phi} \Bigr ] \\
\ln \Bigl (\frac{1 }{1-e^{-i \phi}/a} \Bigr ) E(x+y,z+t,\phi)
\label{expecteta}
\end{eqnarray}
where $W_+$ is given by $(\ref{finwplus})$, $E(x+y,z+t,\phi)$ is defined by $(\ref{EXYZT})$
 and $ \varepsilon $ is
given in $(\ref{epsilon})$.
We would like to stress that $<\eta>$ does not depend on  $a$ if it obeys
unequality $(\ref{agreat1})$. It is convenient for numerical calculations 
to put $a$ close to the unity 
for a better convergency of integrals. 
To see that $<\eta>$ given by $(\ref{expecteta})$ does not coincide with the physical asymmetry $C$
defined in $(\ref{asymback})$ for the case when the background process contribution is important,
let us decompose $(\ref{expecteta})$ into power series with respect to $z$ and $t$. Considering them
as small parameters and neglecting terms $\sim t^2$, $tz$, $z^2$ and  terms of higher orders 
it is easy to get the expression
\begin{equation}
<\eta>= 
\frac{x-y}{x+y}+2(tx-yz) \Bigl [ \frac{\phi(x+y)+e^{-x-y} } {1-e^{-x-y} }
- \frac{1}{x+y}  \Bigr] +...\;\;\;.
\label{decometa}
\end{equation} 
Expression $(\ref{decometa})$ can be compared with the power series for $C$ 
\begin{equation}
C= 
\frac{x-y}{x+y}+2 \frac{ (tx-yz)}{(x+y)^2}  +...
\label{decomc}
\end{equation} 
which follows from  $(\ref{asymback})$. We see from a comparison of 
$(\ref{decometa})$ and $(\ref{decomc})$ that the difference
\begin{equation}
<\eta>-C= 
2(tx-yz) \Bigl [ \frac{\phi(x+y)+e^{-x-y} } {1-e^{-x-y} }
- \frac{1}{x+y} - \frac{1}{(x+y)^2} \Bigr] +...
\label{decdif01}
\end{equation} 
 is nonzero. Applying $(\ref{asymphi})$ we obtain from $(\ref{decdif01})$ the asymptotic 
expression for the difference $<\eta>-C$ valid at $x+y \gg 1$
\begin{equation}
<\eta>-C=  \frac{4(tx-yz)}{(x+y)^3}+...\;.
\label{asdif01}
\end{equation} 
Formula $(\ref{asdif01})$ becomes especially simple if there is no 
asymmetry for background events. Indeed, if 
 $z=t$ we get instead of $(\ref{asdif01})$
\begin{equation}
\frac{ <\eta>-C}{C}=  \frac{2t}{x+y} \cdot \frac{2}{x+y}...\;\;\;. 
\label{asdif02}
\end{equation} 
Relation $(\ref{asdif02})$ shows that the fractional difference between the expected value of
the random variable $\eta$ and the physical asymmetry $C$ is proportional to two small factors:
the background to signal ratio, $2t/(x+y)$ and the inverse value of the total number 
of events, $(x+y)^{-1}$.

To calculate the variance of the random variable $\eta$ defined by $(\ref{difeta})$ we are 
to calculate $<\eta ^2>$ which (by definition) is given by the formula
 \begin{eqnarray}
\nonumber
<\eta ^2>=< \Bigl ( \frac{p-m-n+k}{p+n-m-k}\Bigr )^2>=\\
= \frac{\varepsilon}{1-W_-}
  \sum_{p+n>m+k} \Bigl (\frac{x^p\; y^n\; z^m\; t^k}{p!\; n!\; m!\; k!}\Bigr )
\Bigl ( \frac{p-n-m+k}{p+n-m-k}\Bigr ) ^2 \;.         
\label{seceta01}
\end{eqnarray}
We make use of formulas $(\ref{jback})$ with $j=2$, $(\ref{bessel01})$, $(\ref{bessel02})$ 
to calculate sums over $p$, $n$, $m$, $k$ in $(\ref{seceta01})$ as it has been demonstrated
when obtaining the expected value of $\eta$. The final formula for $<\eta ^2>$ reads
 \begin{eqnarray}
\nonumber
<\eta ^2>= 
\frac{ \varepsilon }{2 \pi W_+} \int_0^{2 \pi} d \phi
\int_0^{ \infty} \frac{ d \alpha \alpha e^{- \alpha } E(x+y,z+t,\phi)}
{1-e^{-\alpha}e^{-i \phi}/a }\\
\nonumber
\Bigl [ (x-y)^2e^{-\alpha}+
 (x+y)+ \frac{1}{a^2}(z+t)e^{-2 i \phi} +\\
\nonumber
\frac{2}{a}(x-y)(t-z)e^{-i \phi}+ \frac{1}{a^3}(t-z)^2e^{-3i \phi}\Bigr ] + \\
\frac{ \varepsilon (x-y)^2}{ W_+} \sqrt{\frac{z+t}{x+y}} 
I_1 \Bigl (2\sqrt{(x+y)(z+t)}\;\Bigr )
\label{squareta}
\end{eqnarray}  
where $W_+$, $E(x+y,z+t,\phi)$ and  $\varepsilon$ are defined in  $(\ref{finwplus})$, $(\ref{EXYZT})$ 
and  $(\ref{epsilon})$, 
respectively.  To compute the variance of $\eta$ for region 
$(\ref{reg02})$ we are to put 
$(\ref{squareta})$ and $(\ref{expecteta})$ into the well known formula
\begin{equation}
\delta \eta  ^2=<\eta ^2>-<\eta > ^2 \;. 
\label{dispeta}
\end{equation}  

Considering again $z$ and $t$ as small parameters, decomposing 
$\delta \eta ^2$ into power series
with respect to $z$, $t$ and retaining terms $\sim z^0$, $z^1$, $t^0$, $t^1$ 
only we get for region $(\ref{reg02})$
 \begin{eqnarray}
\nonumber 
\delta \eta  ^2=\Bigl [1-\Bigl (\frac{x-y}{x+y}\Bigr )^2 \Bigr ]
\Bigl \{ \frac{\phi(x+y)}{1-e^{-x-y}}+(z+t)\Bigl [1-\frac{\phi(x+y)}{1-e^{-x-y}}\\
\nonumber 
-(x+y)\phi(x+y) [1- (e^{x+y}-1)^{-2}]-\frac{x+y}{e^{x+y}-1} \Bigr ]  \Bigr \} +\\
\nonumber
4\frac{tx+yz}{1-e^{-x-y}}\chi(x+y)\\
+4\frac{(tx-yz)(x-y)}{x+y}
\Bigl [\frac{1}{x+y}-\frac{\phi(x+y)+e^{-x-y} }{1-e^{-x-y} }\Bigr ]
\label{decdisp01}
\end{eqnarray}  
where $\phi(z)$ is given by $(\ref{intrephi2})$ and the formula for $\chi(z)$ reads 
\begin{eqnarray}
\nonumber
\chi(z)=e^{-z} \int_0^{z}\frac{e^{t}}{t}\phi(t) dt= e^{-z}  
\sum_{m=1}^{\infty}\frac{z^{m}}{m^2\;m!}= \\
\frac{z}{2}\int_0^{1} \ln ^2(1-t)e^{-zt} dt \;. 
\label{defchi}
\end{eqnarray} 
The behaviour of the function $\chi(z)$ is presented in Fig.~1 with the dashed curve.
To obtain the asymptotic behaviour of $\delta \eta  ^2$ at large $x+y$,
 we are to use 
formula 
$(\ref{asymphi})$ and the asymptotic power series for 
$\chi(z)$ at $z \rightarrow \infty$
\begin{equation}
\chi(z)= \sum_{m=2}^{\infty}\frac{g_{m}}{z^m}= \frac{1}{z^2}+\frac{3}{z^3}+\frac{11}{z^4}
+\frac{50}{z^5}+...
 \label{asympchi}
\end{equation} 
where the coefficients $g_{m}$ in $(\ref{asympchi})$ obey the recurrent relation
\begin{equation}
g_{m}=(m-1)g_{m-1}+(m-2)!
 \label{recgm}
\end{equation} 
with $g_2=1$. Substitution of $(\ref{asymphi})$ and $(\ref{asympchi})$ 
into $(\ref{decdisp01})$ 
leads to the relation of interest valid at $x+y \rightarrow \infty$ 
\begin{eqnarray}
\nonumber
\delta \eta  ^2=\Bigl [1-\Bigl (\frac{x-y}{x+y}\Bigr )^2 \Bigr ]
\Bigl [\frac{1}{x+y}+\frac{1}{(x+y)^2} \Bigr ] \\+
4\frac{t+z}{(x+y)^2}-4\frac{ty^2+zx^2}{(x+y)^4}+...\;\;.
\label{assigma01}
\end{eqnarray} 

If we consider the expected value of $\eta$ in region $(\ref{reg01})$ we are to make use of the
relation
\begin{equation}
<\eta>(1-W_0)=U_p-U_n-U_m+U_k +V_p-V_n-V_m+V_k\;, 
\label{etaexp01}
\end{equation} 
where $W_0$ denotes a total probability of events for which $p+n=m+k$. The formula 
for $W_0$ 
follows immediately from  $(\ref{wminus})$ if we keep the term with $r=0$
in the sum and  we have as a result  
\begin{equation}
W_0= \varepsilon I_0\Bigl (2\sqrt{ (x+y)(z+t)} \;\Bigr )= 
  \frac{\varepsilon }{2 \pi } \int_0^{2 \pi} d \phi 
E(x+y,z+t,\phi) d \phi \;.
\label{w0}
\end{equation} 
In $(\ref{etaexp01})$ $U_s$ denote quantities ($s=p,\;n,\;m,\;k$)  which have been 
calculated above and $V_s$ can be defined by relation  $(\ref{us})$ in which
we are to sum over all $p,\;n,\;m,\;k$ in the region
\begin{equation}
p+n<m+k\;. 
\label{reg03}
\end{equation} 
It is easy to check that $V_p$ coincides with $(-U_k)$ in which we are to make substitutions 
$x \leftrightarrow t,\; y \leftrightarrow z$. The expression for
$V_k$ can be obtained from $(-U_p)$ after the same substitutions. 
Applying  the substitutions $x \leftrightarrow y,\; 
t \leftrightarrow z$ one gets the formulas for 
$V_n$ and $V_m$ from the relations for $V_p$ and $V_k$, respectively.
Putting expressions for $W_0$, $U_s$, $V_s$ ($s=p,\;n,\;m,\;k$) into $(\ref{etaexp01})$ we get 
the final formula for the expected value of $\eta$ in region $(\ref{reg01})$ 
\begin{eqnarray}
\nonumber
<\eta>= \\
\nonumber
\frac{ \varepsilon }{ 2\pi } \int_0^{2 \pi} d \phi  \Bigl \{
\Bigl [a (x-y)e^{i \phi}+\frac{t-z}{a}e^{-i \phi} \Bigr ] 
E(x+y,z+t,\phi)\\
\nonumber 
-\Bigl [a(t-z)e^{i \phi}+\frac{x-y}{a}e^{-i \phi} \Bigr ] 
E(z+t,x+y,\phi)\Bigr \}\\
\ln \Bigl (\frac{1 }{1- e^{-i \phi}/a} \Bigr )  
\Bigl [1- \varepsilon I_0(2\sqrt{(x+y)(z+t)})\Bigr ]^{-1} 
\label{etaexpect02}
\end{eqnarray}
where $\varepsilon$ is defined in $(\ref{epsilon})$ and $E(z+t,x+y,\phi)$ is equal to
\begin{eqnarray}\nonumber
E(z+t,x+y,\phi)=\exp \{ a(t+z)e^{i \phi}+ \frac{1}{a}(x+y)e^{-i \phi} \} 
\end{eqnarray}
in accordance with $(\ref{EXYZT})$.

Decomposing  $(\ref{etaexpect02})$ into power series with respect to $z$ and $t$ we get
\begin{eqnarray}
\nonumber
<\eta>= 
\frac{x-y}{x+y}\\
+\frac{ 2(tx-yz)}{1-e^{-x-y}} \Bigl [ \phi(x+y)+e^{-x-y} 
- \frac{1}{x+y}  \Bigr] +...\;.
\label{decometa02}
\end{eqnarray} 
Remembering $(\ref{decomc})$  we get easily for the difference $\eta -C$ the relation
\begin{eqnarray}
\nonumber
<\eta>-C= \\
2(tx-yz) \Bigl [ \frac{\phi(x+y)+e^{-x-y} -(x+y)^{-1}} {1-e^{-x-y} }
- \frac{1}{(x+y)^2} \Bigr] +...
\label{decdif02}
\end{eqnarray} 
instead of $(\ref{decdif01})$. Formula $(\ref{decdif02})$ shows that $<\eta>$ does not 
coincide with the physical asymmetry $C$ in region $(\ref{reg01})$  if the background 
contribution is not negligible. This is true in spite of  applying the subtraction 
procedure: we consider $p-m$ and 
$n-k$ instead of $p$ and $n$ in $(\ref{difeta})$. It is easy to check that the first term 
in the asymptotic formula for $<\eta>-C$ obtained from  $(\ref{decdif02})$ coincides with
$(\ref{asdif01})$. If one puts $x$, $y$, $z$, $t$ equal to the observed numbers of the experimental
 events $x=p$, $y=n$, $z=m$, $t=k$, then formulas $(\ref{decdif01})$ and $(\ref{decdif02})$ 
can be used to estimate the systematic uncertainty due to the difference between 
the  expected value of the random variable $\eta$ and the physical asymmetry $C$. 
They can be improved if we make use of the precise expression for $C$  and $<\eta>$ given by 
$(\ref{asymback})$ and $(\ref{expecteta})$,  $(\ref{etaexpect02})$, respectively.

To compute the second moment of $\eta$ in region $(\ref{reg01})$ we may use formula 
  $(\ref{seceta01})$ if we add in the sum in the right hand side 
a contribution of events with $p+n<m+k$ 
and put $W_0$ instead of $W_-$. This contribution is described 
by formula   $(\ref{squareta})$ in which
we should  use the substitution $x \leftrightarrow t$, 
$y \leftrightarrow z$, and $W_+ \rightarrow 1-W_0$. 
The expression for $<\eta ^2>$ looks like
 \begin{eqnarray}
\nonumber
<\eta ^2>= \\
\nonumber 
\frac{ \varepsilon }{2 \pi (1-W_0)} \int_0^{2 \pi} d \phi
\int_0^{ \infty}\Bigl (   \frac{ d \alpha \alpha e^{- \alpha }}
{1-e^{-\alpha}e^{-i \phi}/a} \Bigr )
E(x+y,z+t,\phi) \\
\nonumber
\Bigl [ (x-y)^2e^{-\alpha}+
 (x+y)+ \frac{1}{a^2}(z+t)e^{-2 i \phi} \\
\nonumber
+\frac{2}{a}(x-y)(t-z)e^{-i \phi}+ \frac{1}{a^3}(t-z)^2e^{-3i \phi}\Bigr ] +\\
\nonumber
\frac{ \varepsilon }{2 \pi (1-W_0)} \int_0^{2 \pi} d \phi
\int_0^{ \infty} \Bigl (  \frac{ d \alpha \alpha e^{- \alpha }}
{1-e^{-\alpha}e^{-i \phi}/a} \Bigr )
E(t+z,x+y, \phi)\\
\nonumber
\Bigl [ (t-z)^2e^{-\alpha}+
 (t+z)+ \frac{1}{a^2}(x+y)e^{-2 i \phi} \\
\nonumber
+\frac{2}{a}(x-y)(t-z)e^{-i \phi}+ \frac{1}{a^3}(x-y)^2e^{-3i \phi}\Bigr ]+\\ 
\nonumber
\frac{ \varepsilon }{1- W_0} I_1 \Bigl (2\sqrt{(x+y)(z+t)}\;\Bigr )  
 \Bigl [(x-y)^2 \sqrt{\frac{z+t}{x+y}} \\
+(t-z)^2\sqrt{\frac{x+y}{t+z} } \Bigr ]
\label{squareta2}
\end{eqnarray}  
with $W_0$ defined in $(\ref{w0})$. Putting $(\ref{squareta2})$ and  $(\ref{etaexpect02})$ into 
$(\ref{dispeta})$ we come to the formula of interest for the variance of $\eta$ in the region
$(\ref{reg01})$. A decomposition of the formula for the variance into power series  up to terms 
$z$ and $t$ gives the relation
 \begin{eqnarray}
\nonumber 
\delta \eta  ^2=\Bigl [1-\Bigl (\frac{x-y}{x+y}\Bigr )^2 \Bigr ]
\Bigl \{ \frac{\phi(x+y)}{1-e^{-x-y}}+\frac{(z+t)}{1-e^{-x-y}} 
\Bigl [1-\frac{\phi(x+y)}{1-e^{-x-y}}\\
\nonumber 
-(x+y)\phi(x+y) \frac{1- 2e^{-x-y}}{1-e^{-x-y}}-(x+y)e^{-x-y} \Bigr ]  \Bigr \} \\
\nonumber
+\frac{4(tx+yz)}{1-e^{-x-y}}\chi(x+y)\\
+\frac{4(tx-yz)(x-y)}{(x+y)(1-e^{-x-y})}
\Bigl [\frac{1}{x+y}-\phi(x+y)-e^{-x-y}  \Bigr ]
\label{decdisp02}
\end{eqnarray}  
with $\phi(z)$ and $\chi(z)$ defined in  $(\ref{intrephi2})$ 
and  $(\ref{defchi})$, respectively.
The asymptotics of $(\ref{decdisp02})$ at $x+y \gg 1$ is given  
by relation $(\ref{assigma01})$.

The most important limit of the obtained formulas corresponds to the case when $x+y+z+t$ go to
infinity but the ratio $Q=(z+t)/(x+y)$ is less than the unity. For this case the contribution of the
region
$p+n<m+k$ is exponentially small. Indeed, making use of the well known asymptotic formula for the
Bessel functions \cite{Whitt}
\begin{equation}
I_n(z)=\frac{1}{\sqrt{2 \pi z}}e^z
\label{asympinz}
\end{equation}
valid at $z \rightarrow \infty$ we get from $(\ref{wminus})$ the relation
 \begin{eqnarray}
\nonumber
W_-= \frac{1}{\sqrt{4 \pi}} \sum_{m=0}^{\infty}\Bigl (\frac{z+t}{x+y} \Bigr )^{m/2} \Bigl \{
\frac{e^{-x-y-z-t}}{[(x+y)(z+t)]^{1/4}}\Bigr \} \\
\nonumber
 \exp \{ 2 \sqrt{(x+y)(z+t)} \}= \\
\frac{1}{\sqrt{4 \pi}} \frac{\sqrt{x+y}\;\exp \{ - (\sqrt{x+y}-\sqrt{z+t})^2 \}}
{(\sqrt{x+y}-\sqrt{t+z})\;[(x+y)(z+t)]^{1/4}}
\label{wmintozer}
\end{eqnarray}
which shows that $W_- \rightarrow 0$ exponentially if 
\begin{equation}
 (\sqrt{x+y}-\sqrt{z+t})^2  \gg 1\;.
\label{exponcond}
\end{equation}
Ignoring such  exponentially small corrections  we may obtain the asymptotic formula for $<\eta>$
 which reads
 \begin{eqnarray}
\nonumber
<\eta>= \Bigl (\frac{x-y+t-z}{x+y-t-z}\Bigr ) \Bigl [1+\frac{x+y+z+t}{(x+y-z-t)^2}\\ 
\nonumber
+\frac{3(x+y+z+t)^2}{(x+y-z-t)^4}+ 
\frac{1}{(x+y-z-t)^2}+... \Bigr ] \\-
\frac{x-y-t+z}{(x+y-t-z)^2} \Bigl [1+\frac{3(x+y+z+t)}{(x+y-z-t)^2}+... \Bigr ]\;.
\label{etaxyztinf}
\end{eqnarray}
Comparison of $(\ref{etaxyztinf})$ with $(\ref{asymback})$ shows that $<\eta>$ coincides with the
physical asymmetry $C$ if 
\begin{equation}
\frac{x+y+z+t}{(x+y-z-t)^2} \ll 1\;.
\label{condeta}
\end{equation}  
It is easy to see that both conditions $(\ref{exponcond})$ and $(\ref{condeta})$ are valid if 
$x+y+z+t \gg 1$ and $Q=(z+t)/(x+y)<1$.
If conditions $(\ref{exponcond})$ and $(\ref{condeta})$ are fulfilled the variance of the random
variable $\eta$ looks like
\begin{eqnarray}
\nonumber
\delta \eta ^2=\frac{x+y+z+t}{(x+y-z-t)^2} \Bigl [1+\frac{(x-y-z+t)^2}{(x+y-z-t)^2} \Bigr ]
\\-2\frac{(x-y)^2-(z-t)^2}{(x+y-z-t)^3}\;.
\label{varsigma}
\end{eqnarray}
We have retained in $(\ref{varsigma})$ the greatest terms in the asymptotic power series only.
Formulas  $(\ref{etaxyztinf})$ and $(\ref{varsigma})$  correspond to the
Gaussian distribution for
the random variable $\eta$. In particular, relation $(\ref{varsigma})$ can be obtained from the
widely used formula
\begin{equation}
\delta \eta ^2=\Bigl (\frac{\partial \eta}{\partial p} \Bigr ) ^2 \delta p ^2+
\Bigl (\frac{\partial \eta}{\partial n} \Bigr ) ^2 \delta n ^2+
\Bigl (\frac{\partial \eta}{\partial m} \Bigr ) ^2 \delta m ^2+
\Bigl (\frac{\partial \eta}{\partial k} \Bigr ) ^2 \delta k ^2
\label{partvar}
\end{equation}
where the dependence of $\eta$ on $p$, $n$, $m$, $k$ is given by $(\ref{difeta})$ and $\delta p
^2=p$,
$\delta n ^2=n$, $\delta m ^2=m$, $\delta k ^2=k$ in accordance with the Poisson distributions
$(\ref{poiss})$, $(\ref{backpoiss})$ for the random variables $p$, $n$, $m$, $k$. If one can
neglect the background contribution ($z \rightarrow 0$, $t \rightarrow 0$) formula
$(\ref{varsigma})$ reduces to $(\ref{asymsigma})$.
 Formulas $(\ref{etaxyztinf})$ and $(\ref{varsigma})$ are obtained in
Appendix.

Figures~4 and 5 show the dependence of the ratio $f=$
\linebreak $<\eta>/C$ on $\lambda$ where 
$C$ is defined by $(\ref{asymback})$ and the relation between $x$, $y$, $z$, $t$  and $\lambda$
looks like 
\begin{equation}
x=x_0 \lambda\;,\;\;\;y=y_0 \lambda\;,\;\;\;z=z_0 \lambda\;,\;\;\;t=t_0 \lambda
\label{x0y0z0t0}
\end{equation}
 the parameters $x_0$, $y_0$, $z_0$, $t_0$
being chosen in such a way that $x_0+y_0+z_0+t_0=1$. The solid lines are
calculated in  region $(\ref{reg02})$ with the aid of formula $(\ref{expecteta})$
and dashed curves are
obtained for region $(\ref{reg01})$ using relation $(\ref{etaexpect02})$. The dash-dotted curves
are computed with the asymptotic formula $(\ref{etaxyztinf})$. Figure~4 shows that all the ratios
go to the unity with an increase of the total statistics but the deviation from the unity increases
if the mean value of the background to signal ratio b/s=(z+t)/(x+y-z-t) increases that follows
from
a comparison of Fig.~4a, Fig.~4b and Fig.~4c for which $b/s=12.5\%,\;33\%,\;75\%$, respectively.
We see also that the higher the background to signal ratio the slower a value of $<\eta>/C$ tends
to the unity with  increasing  $\lambda$.
The deviation from the unity for the curves presented
in Fig.~4d is larger than for those of Fig.~4c though the background to signal ratio is the same for
these two cases. It is due to the fact that the background asymmetry is equal to zero ($z=t$) for
the curves shown in Fig.~4c while it has the opposite sign than the measured asymmetry
($z-t<0$, $C>0$) for the case presented in Fig.~4d. The behaviour of the ratio
$f=<\eta>/C$ at small  $C$ ($C \sim 1\%$) is presented in Fig.~5. Curves in Figs.~5a, 5b, 5c
correspond to the physical asymmetry $C=0.025$ and $b/s=12.5\%$ but the background events have
zero, negative and
positive asymmetry, respectively. We see that the greatest deviation of the curves
from the unity corresponds to the case when
$C$ and the background asymmetry have opposite signs. The curves in Fig.~5d are calculated for the
large background contribution ($b/s=75\%$) nevertheless the maximum deviation of the curves in
this case is 
even smaller than in Figs.~5b, 5c. 
This example shows that the asymmetry of the low background
 may lead to higher difference between $<\eta>$ and $C$ than a high background without the
asymmetry. This is true for low statistics only. We see from Figs.~4 and 5 that at large
$\lambda$
the deviation of $f$ from the unity is greater for the high background than for 
the low
background.
It is easy to see that the
deviations
of all curves from the unity is very appreciable. Figures~4 and 5 illustrate a usefulness of
formulas $(\ref{expecteta})$, $(\ref{etaexpect02})$ which can be applied for  estimating
the difference between $<\eta>$ and the physical
asymmetry $C$.
We see from Figs.~4 and 5 that the difference
between formulas $(\ref{expecteta})$ and $(\ref{etaexpect02})$ reaches about few per cents when
$\lambda=x+y+z+t$ is about few dozens. It is easy to see also from a comparison of the curves
presented in  Figs.~4 and 5 that the asymptotic relation $(\ref{etaxyztinf})$ predicts $<\eta>$
with rather high accuracy for $\lambda \geq 20 -30$ even for the essential contribution
of the background events when the deviation of $<\eta>$ from $C$
is appreciable ($\sim 20\%$).

\section{Conclusions}

We have obtained the precise formulas for the expected values and the
variances of the random variables $\xi$ and $\eta$ which correspond to the asymmetry of
some reaction when the background process contribution is negligible and
appreciable, respectively. Introducing the conditional probability distribution
we have established that the expected value of $\xi$ is equal to the physical  
asymmetry $C$  (see $(\ref{mathfin})$) if there is no background process
contribution to the reaction under study. We have obtained precise formula
$(\ref{uncerfin})$ for the variance of the random variables $\xi$ which is valid if
one may ignore the background  contribution. Relation $(\ref{uncerfin})$
shows that the variance is finite, and hence $\xi$ has nothing common with the property
of the random variable with the 
\linebreak
\begin{center}
\mbox{\epsfig{file=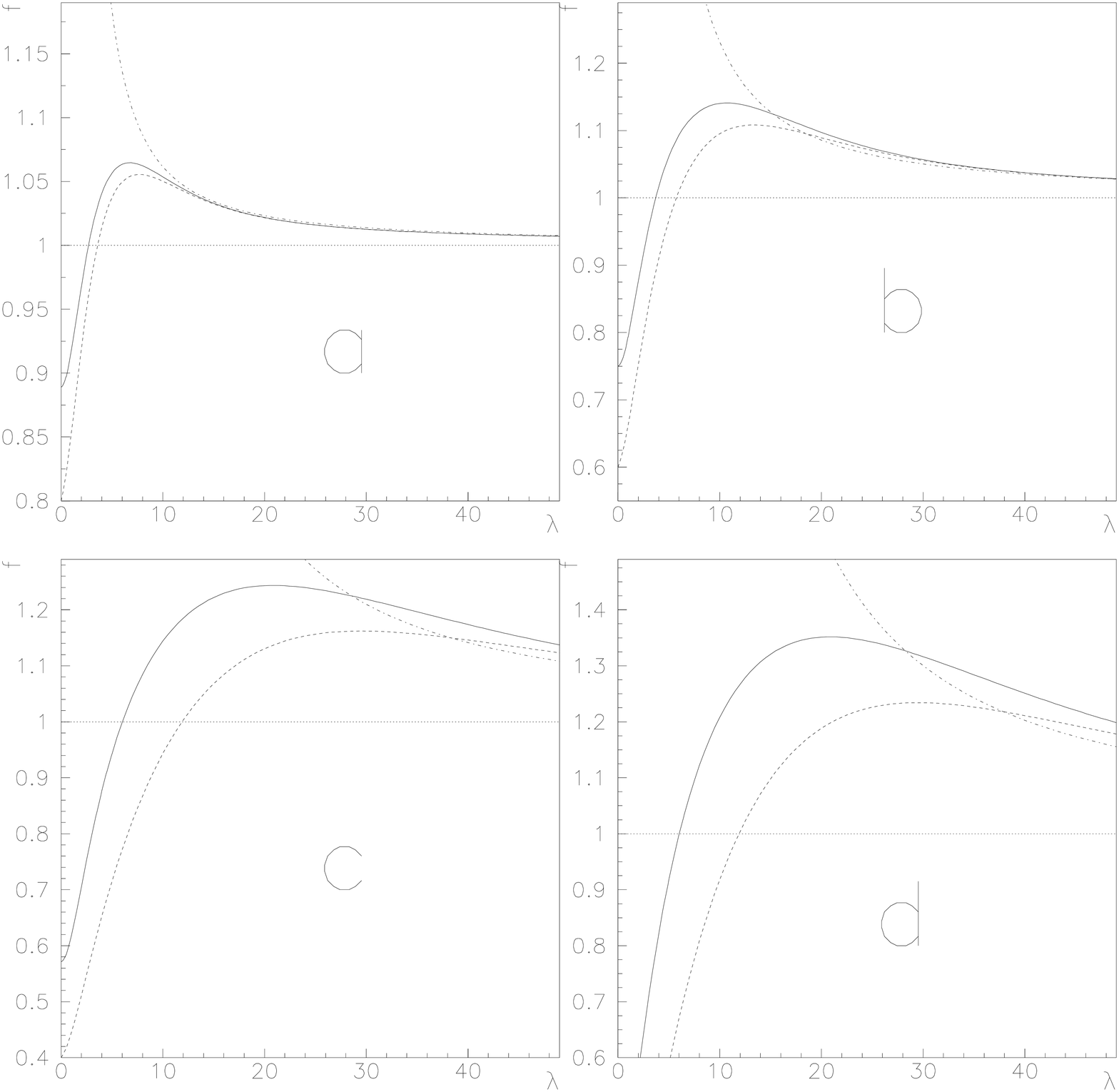,height=16cm,width=15.5cm}}
\noindent
\small
\begin{minipage}{15cm}
\vspace{8mm}
{\sf Fig.~4:}   
Dependence of ratios $<\eta>/C$ on $\lambda$.
Solid, dashed and dash-dotted curves are calculated using $(\ref{expecteta})$,
$(\ref{etaexpect02})$ and $(\ref{etaxyztinf})$,
respectively. 
Parameters are: a) $x_0=0.55$, $y_0=0.35$, $z_0=0.05$, $t_0=0.05$, 
$C=0.25$, $b/s=0.125$
b) $x_0=0.5$, $y_0=0.3$, $z_0=0.1$, $t_0=0.1$, $C=1/3$, $b/s=1/3$
c) $x_0=0.45$, $y_0=0.25$, $z_0=0.15$, $t_0=0.15$, $C=0.5$, $b/s=0.75$
d) $x_0=0.45$, $y_0=0.25$, $z_0=0.1$, $t_0=0.2$, $C=0.75$, $b/s=0.75$. 
Parameters $x_0$, $y_0$, $z_0$, $t_0$ and $\lambda$ are defined in $(\ref{x0y0z0t0})$, $C$ is given
by $(\ref{asymback})$ and $b/s=(z+t)/(x+y-z-t)$ denotes the background to signal ratio.
\end{minipage}
\end{center}  
%%%%\vspace{0.3cm}
\normalsize
\begin{center}
\mbox{\epsfig{file=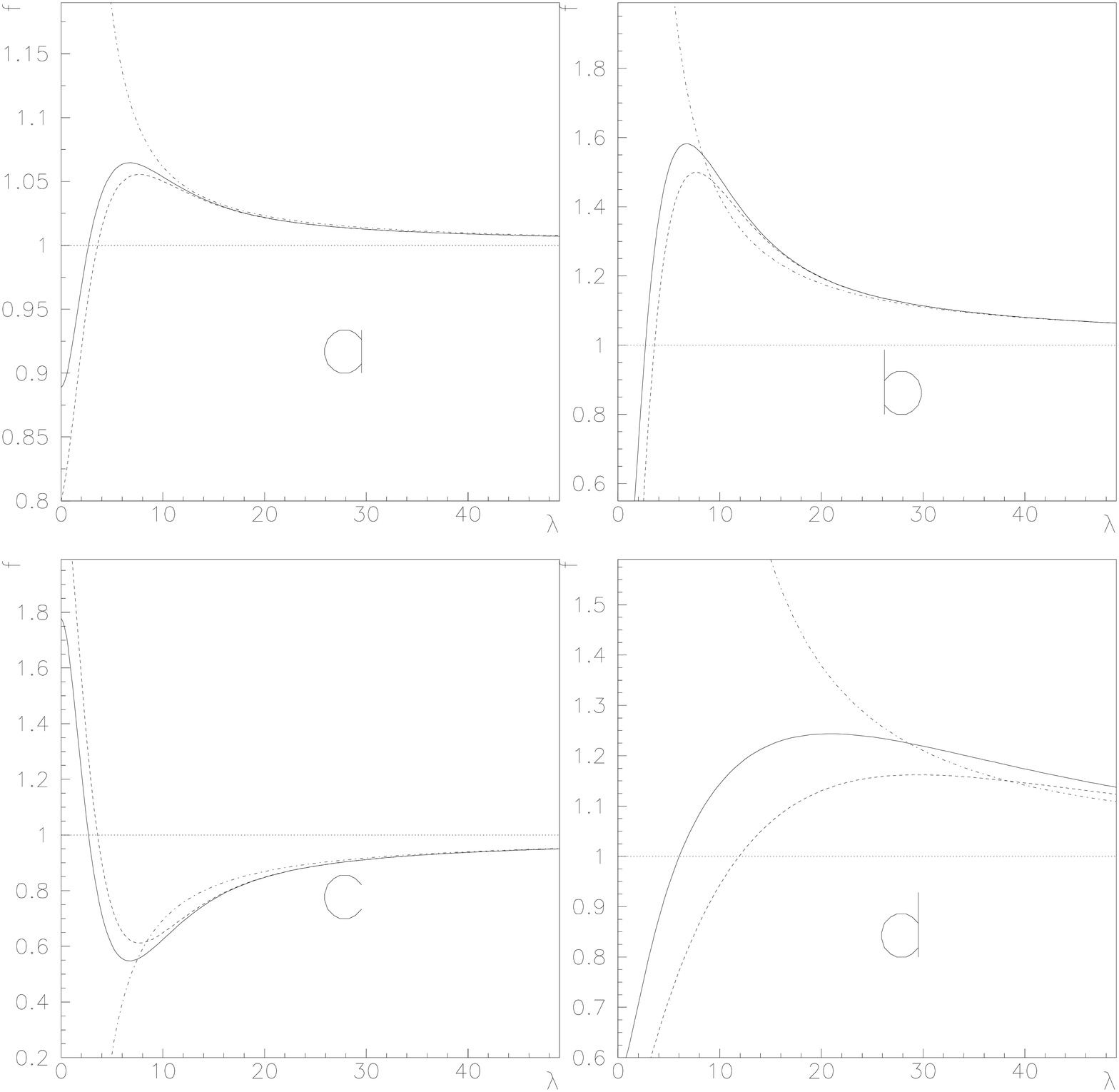,height=16cm,width=15.5cm}}
%angle=270}}   XS
\noindent
\small
\begin{minipage}{15cm}
\vspace{8mm}
{\sf Fig.~5:}
Dependence of ratios $<\eta>/C$ on $\lambda$.
Solid, dashed and dash-dotted curves are calculated using $(\ref{expecteta})$,   
$(\ref{etaexpect02})$ and $(\ref{etaxyztinf})$,
respectively. 
Parameters are: a) $x_0=0.46$, $y_0=0.44$, $z_0=0.05$, $t_0=0.05$,
$C=0.025$, $b/s=0.125$
b) $x_0=0.45$, $y_0=0.45$, $z_0=0.04$, $t_0=0.06$, $C=0.025$, $b/s=0.125$
c) $x_0=0.47$, $y_0=0.43$, $z_0=0.06$, $t_0=0.04$, $C=0.025$, $b/s=0.125$
d) $x_0=0.355$, $y_0=0.345$, $z_0=0.15$, $t_0=0.15$, $C=0.025$, $b/s=0.75$.
All notations are the same as in Fig.~4.
\end{minipage}
\end{center}  
\vspace{0.3cm}
\normalsize
Cauchy density function. This is the result of 
defining $\xi$ in the region $p+n >0$ and making use of the conditional 
probability distribution. Formula $(\ref{uncerfin})$ predicts that the variance 
 vanishes if the physical asymmetry squared tends to the unity. For a large
number of experimental events $N^{exp}$ the standard deviation 
$\delta \xi$ goes to zero as $ \sqrt{1/N^{exp}}$ according to 
$(\ref{asymsigma})$. The finiteness of the variance of $\xi$ 
 allows to utilize all the experimental statistics accumulated in many  bins
(for which  expected values of the random variables describing the measured 
asymmetry are equal to each other)  to

get a higher statistical accuracy of the asymmetry. We may
consider the random variable 
$\zeta$ defined by  $(\ref{defzet})$ which has the expected value equal 
to the studied asymmetry but with the variance  smaller than 
the variance $\delta \zeta  _j ^2$ which corresponds to the statistical uncertainty 
of the asymmetry obtained from the $j$th bin. 

As  follows from
$(\ref{decdif01})$ and $(\ref{decdif02})$  the expected value of $\eta$ both
in region $(\ref{reg01})$ and $(\ref{reg02})$ does not coincide with the physical
asymmetry $C$ (defined in  $(\ref{asymback})$) for the  process under 
investigation if the background  contribution is not negligible. Relations
 $(\ref{expecteta})$ and  $(\ref{etaexpect02})$ give the precise expressions for
the expected values of the random variable $\eta$ obtained with the conditional 
probability distributions defined in regions  $(\ref{reg02})$ 
and  $(\ref{reg01})$, respectively. Exact formulas $(\ref{dispeta})$,
$(\ref{squareta})$ and $(\ref{expecteta})$ give the expression for the variance of 
$\eta$ in region $(\ref{reg02})$. Precise relations $(\ref{dispeta})$, 
$(\ref{squareta2})$ and $(\ref{etaexpect02})$ are to be used for computing
the variance of the random variable $\eta$ defined in region $(\ref{reg01})$. 
We conclude from these relations that the variance is finite in both 
cases under discussion. This important result means that we may use the statistics 
of the experimental events accumulated in many bins to reduce the statistical 
uncertainty of the measured asymmetry. We would like to stress that, nevertheless,
the obtained
asymmetry has a systematic uncertainty since the expected value of $\eta$ deviates
from the true physical asymmetry $C$ in every bin. Formulas  $(\ref{decdif01})$ and 
$(\ref{decdif02})$
can be applied to estimate  the systematic errors due to the background
contribution.  
The asymptotic formulas $(\ref{etaxyztinf})$ and $(\ref{varsigma})$ for $<\eta>$ and
$\delta \eta  ^2$ have been obtained for  
$x+y+z+t \gg 1$ and
$(z+t)/(x+y)<1$.
This high statistics limit  shows that $<\eta>$
coincides with the physical asymmetry $C$ and the standard deviation 
%%%\newline
$\delta \eta  \sim
\sqrt{x+y+z+t}/ 
(x+y-z-t) \ll 1$
%%%\newline
if conditions $(\ref{exponcond})$ and $(\ref{condeta})$
are fulfilled. 

\section*{Acknowledgement}
I am grateful to A.~V.~Kravtsov and S.~G.~Sherman for the help which improved
this manuscript.

\section{Appendix}

Let us consider the asymptotic behaviour of $<\eta>$ and $\delta \eta  ^2$ 
when $x+y+z+t \rightarrow \infty$
but the ratio $Q=(z+t)/(x+y) < 1$ . As has been mentioned the final results do not depend on 
the value of $a$ but it is convenient to consider the limit of
$(\ref{finwplus})$ and $(\ref{expecteta})$ at $a \rightarrow 1$ ($a>1$). 
Since the integrands in $(\ref{finwplus})$ and $(\ref{expecteta})$ at $\phi=\theta$ and $\phi=2
\pi-\theta$ 
are complex conjugated functions  we may
integrate from $0$ to $\pi$ considering the real part of the integrals only. It is easy to see that
the dominant contribution to the integrals in $(\ref{finwplus})$ and $(\ref{expecteta})$ comes
from the
regions $0\leq \phi \leq \beta$ and $2\pi-\beta \leq \phi \leq 2\pi$ with $\beta ^2 \sim (x+y+z+t)
^{-1} \ll 1$ if $x+y+z+t$ is very large. Decomposing $e^{\pm i \phi}$ into power series up to terms
$\phi ^2$ we get
 \begin{eqnarray}
\nonumber
W_+ \approx 1-\frac{\epsilon}{\pi}  \int_0^{\pi} d \phi
\mbox{Re} \Bigl \{
\Bigl [\frac{1}{1-Q}-\frac{i \phi Q}{(1-Q)^2}-
\frac{ \phi ^2 Q(1+Q)}{2(1-Q)^3} \Bigr ]\\
\exp \{(x+y+z+t)(1- \frac{\phi ^2}{2})+
i \phi (x+y-z-t) \}  \Bigr \}. 
\label{decwplus}   
\end{eqnarray}
Neglecting exponentially small corrections it is possible to integrate in 
$(\ref{decwplus})$ within the limits $0$ and $\infty$. Making use of 
the following asymptotic formulas valid for
positive and large $u$ and $v$
 \begin{eqnarray}
\nonumber
K_0(u,v)=\int_0^{\infty}\exp \{-\frac{1}{2}u t^2+ivt\}dt \\
\nonumber
=\frac{i}{v}\Bigl [1+\frac{u}{v^2}+\frac{3u^2}{v^4}
+O(\frac{u^3}{v^6})\Bigr ]\;,\\
\nonumber
K_1(u,v) =\int_0^{\infty}i t \exp \{- \frac{1}{2}u t^2+ivt\}dt =\\
\nonumber
-\frac{i}{v^2}\Bigl [1+\frac{3u}{v^2}   
+O(\frac{u^2}{v^4})\Bigr ]\;,\\
K_2(u,v) =\int_0^{\infty}(i t)^2 \exp \{- \frac{1}{2} u t^2+ivt\}dt =
\frac{2i}{v^3}\Bigl [1+                  
O(\frac{u}{v^2})\Bigr ]
\label{k012}
\end{eqnarray}
(which can be easily derived by integration by parts) and putting $(\ref{k012})$ into
$(\ref{decwplus})$
we get that for large $x+y+z+t$ 
\begin{equation}
W_+=1
\label{wplus01}
\end{equation}
since all the integrals in $(\ref{k012})$ have no real parts. The corrections to formula
$(\ref{wplus01})$ are exponentially small in accordance with $(\ref{wmintozer})$.

Since a behaviour of $\ln (1-e^{-i \phi})$ at small $\phi$ looks like
\begin{equation}
\ln (1-e^{-i \phi})= i \frac{\pi}{2}+\ln \phi -i \frac{\phi}{2}- \frac{\phi ^2}{24}+ ...
\label{declog}
\end{equation}
we need the asymptotics at large positive $u$ and $v$ for  integrals containing logarithms
 \begin{eqnarray}
\nonumber
G_0(u,v)=\int_0^{\infty} \ln t \exp \{-\frac{1}{2}u t^2+ivt\}dt = \\
\nonumber
 \frac{i}{v} \psi(1)+\frac{iu}{v^3}\psi(3)+\frac{3iu^2}{v^5}\psi(5)\\
\nonumber
+ \Bigl[\frac{i \pi}{2}-\ln v \Bigr ] K_0(u,v) +O(\frac{u^3}{v^7})\;,\\
\nonumber
G_1(u,v)=\int_0^{\infty} (i t)\ln t \exp \{-\frac{1}{2}u t^2+ivt\}dt = \\ 
\nonumber
 -\frac{i}{v^2} \psi(2)-\frac{3iu}{v^4}\psi(4)
+ \Bigl[\frac{i \pi}{2}-\ln v \Bigr ] K_1(u,v) +O(\frac{u^2}{v^6})\;,\\
\nonumber
G_2(u,v)=\int_0^{\infty} (i t)^2 \ln t \exp \{-\frac{1}{2}u t^2+ivt\}dt = \\
 \frac{2i}{v^3} \psi(3)
+ \Bigl[\frac{i \pi}{2}-\ln v \Bigr ] K_2(u,v) +O(\frac{u}{v^5})
\label{g012}
\end{eqnarray}
with $\psi(z)= \Gamma '(z)/\Gamma (z)$ being  the  digamma
function.
We  shall obtain the expression for $G_0(u,v)$. Other formulas $(\ref{g012})$ can be derived in
an analogous way. To calculate $G_0(u,v)$ we put the Frullani formula
 \begin{equation}
\ln t = \int_0^{\infty} \frac{d \beta}{\beta}[e^{-\beta}-e^{-\beta t}]
\label{frull}
\end{equation}
into the definition of $G_0$ in $(\ref{g012})$ and integrate over $t$
 \begin{eqnarray}
\nonumber
G_0(u,v)=\int_0^{\infty} \frac{d \beta}{\beta} \int_0^{\infty} 
\Bigl [ \exp \{-\frac{1}{2}u t^2+ivt\}e^{-\beta}\\
\nonumber
- \exp \{-\frac{1}{2}u t^2+i(v+i \beta) t\} \Bigr ]dt
= \\
\nonumber
\int_0^{\infty} \frac{d \beta}{\beta} \Bigl [ e^{-\beta}K_0(u,v)-K_0(u,v+i \beta) \Bigr ] \approx \\
\nonumber
\int_0^{\infty} \frac{d \beta}{\beta} \Bigl [ e^{-\beta} \Bigl (
 \frac{i}{v}+\frac{iu}{v^3}+\frac{3iu^2}{v^5} \Bigr ) \\
\nonumber
 -\frac{i}{v+i \beta}-\frac{iu}{(v+i \beta)^3}-\frac{3iu^2}{(v+i \beta)^5}  \Bigr ]= \\
\nonumber
\int_0^{\infty} \frac{d \beta}{\beta} \Bigl \{
 \frac{i}{v}\Bigl [ e^{-\beta}-\frac{1}{1+i \beta/v} \Bigr ]
+\frac{iu}{v^3}\Bigl [ e^{-\beta}-\frac{1}{(1+i \beta/v)^3} \Bigr ]\\
+\frac{3iu^2}{v^5}\Bigl [ e^{-\beta}-\frac{1}{(1+i \beta/v)^5} \Bigr ] \Bigr \}.
\label{frug0}
\end{eqnarray}
One may write the integral
\begin{eqnarray}
\nonumber 
M_n=\int_0^{\infty} \frac{d \beta}{\beta} \Bigl [e^{-\beta}-\frac{1}{(1+i \beta/v)^n} \Bigr ]
\end{eqnarray}
as the sum of the two integrals
\begin{eqnarray}
\nonumber
M_n=\int_0^{\infty} \frac{d \beta}{\beta} \Bigl [e^{-\beta}-\frac{1}{(1+\beta)^n} \Bigr ] \\
+\int_0^{\infty} \frac{d \beta}{\beta} \Bigl [\frac{1}{(1+\beta)^n}-\frac{1}{(1+i \beta/v)^n}
\Bigr]\;.
\label{mn}     
\end{eqnarray}
The first integral in $(\ref{mn})$ is equal to the digamma function 
 due to the well known Dirichlet formula \cite{Whitt}. To calculate
the second integral in $(\ref{mn})$ we write the chain of the equations
 \begin{eqnarray}
\nonumber
\int_0^{\infty} \frac{d \beta}{\beta} \Bigl [\frac{1}{(1+\beta)^n}-\frac{1}{(1+i \beta/v)^n}
\Bigr]\\
\nonumber
=\frac{(-1)^{n-1}}{(n-1)!}\frac{\partial ^{n-1}}{\partial x^{n-1}}
\int_0^{\infty} \frac{d \beta}{\beta} \Bigl [\frac{1}{x+\beta}-\frac{1}{x+i \beta/v}
\Bigr] \Big |_{x=1}\\
\nonumber
=\frac{(-1)^{n-1}}{(n-1)!}\frac{\partial ^{n-1}}{\partial x^{n-1}}
\int_0^{\infty} \frac{d \beta}{x} \Bigl [\frac{i/v}{x+i\beta/v}-\frac{1}{x+\beta}
\Bigr] \Big |_{x=1} \\
=\frac{(-1)^{n-1}}{(n-1)!}\frac{\partial ^{n-1}}{\partial x^{n-1}}
 \Bigl [\frac{1}{x} \ln \Bigl( \frac{i}{v}\Bigr)\Bigr ] \Big |_{x=1} 
=i\frac{\pi}{2} -\ln v.
\label{mnsec}
\end{eqnarray}
Putting the expressions for  both  integrals in $(\ref{mn})$
we get
\begin{eqnarray}
M_n=\int_0^{\infty} \frac{d \beta}{\beta} \Bigl [e^{-\beta}-\frac{1}{(1+i \beta/v)^n} \Bigr ]
=\psi(n)+i\frac{\pi}{2} -\ln v\;.
\label{mnfin}
\end{eqnarray}
Combining $(\ref{mnfin})$ with $(\ref{frug0})$ we get the expression $(\ref{g012})$ for $G_0(u,v)$.

To compute $<\eta>$, we put $(\ref{wplus01})$ and $(\ref{declog})$ in $(\ref{expecteta})$,
decompose $e^{ \pm i \phi}$ into power series,
integrating
within limits $0$ and $\infty$ and making use of $(\ref{k012})$, $(\ref{g012})$ and get
\begin{eqnarray}
\nonumber   
<\eta> \approx \frac{1}{\pi} \mbox{Re} \Bigl \{\int_0^{\infty} d \phi 
\Bigl [i \frac{\pi}{2}+\ln \phi -
 i\frac{\phi}{2}-\frac{\phi ^2}{24} \Bigr ] \\
\nonumber
\{(x-y+t-z)(1- \phi ^2/2)+i \phi(x-y+z-t)\}\\
\nonumber
\exp \{-\frac{1}{2}(x+y+z+t) \phi ^2
+i \phi (x+y-z-t) \} \Bigr \} \approx \\
\nonumber
(x-y+t-z) \Bigl \{\frac{1}{v}+\frac{u}{v^3}+\frac{3u^2}{v^5}+\frac{1}{v^3} \Bigr \}\\
-(x-y+z-t)\Bigl \{\frac{1}{v^2}+\frac{3u}{v^4} \Bigr \}\;.
\label{etauv}
\end{eqnarray}
Formula $(\ref{etauv})$ coincides with   $(\ref{etaxyztinf})$ since $u=x+y+z+t$
and $v=x+y-z-t$. We would like to note that the contributions to the real 
part give the terms $i \pi
/2$ and $\ln \phi$ in the square brackets in $(\ref{etauv})$,  
their contributions being equal to
each other in accordance with $(\ref{k012})$ and $(\ref{g012})$.  

As has been mentioned above the contribution of region 
\linebreak $p+n<m+k$ vanishes in the limit $x+y+z+t
\rightarrow \infty$, $Q=(z+t)/(x+y)<1$. This means that we are to get the same result 
$(\ref{etauv})$ from  $(\ref{etaexpect02})$. Indeed, $\epsilon I_0$ 
in $(\ref{etaexpect02})$ vanishes exponentially
\begin{eqnarray}
\nonumber
\epsilon I_0(2\sqrt{(x+y)(z+t)}) \\
\sim \exp \{-x-y-z-t+2\sqrt{(x+y)(z+t)} \} \rightarrow 0\;.
\label{wzero}
\end{eqnarray}
A comparison of $(\ref{expecteta})$ and $(\ref{etaexpect02})$ shows that the additional terms in 
$(\ref{etaexpect02})$ 
after the substitution $\phi \rightarrow -\phi$ (we remind that $a=1$)
look like (relation $(\ref{wzero})$ is taken into account)
\begin{eqnarray}
\nonumber
\frac{\epsilon}{\pi} \mbox{Re} \int_0^{\infty}[(x-y)e^{i \phi}+(t-z)e^{-i \phi} ]\\
E(x+y,z+t,\phi)\ln (1-e^{i \phi}) d \phi \;.
\label{addterm}
\end{eqnarray}
A comparison of $(\ref{addterm})$ and $(\ref{expecteta})$ shows that the only difference between
them is the difference between the logarithms. The logarithm $\ln (1-e^{i \phi})$ at small $\phi$
looks like
\begin{equation}
\ln (1-e^{i \phi})= -i \frac{\pi}{2}+\ln \phi +i \frac{\phi}{2}- \frac{\phi ^2}{24}+ ...\;.
\label{declogpl}
\end{equation}
As has been explained above only the terms $i \pi/2$ and $\ln \phi$ in the decomposition of the
logarithm $\ln (1-e^{i \phi})$ contribute to the real part of the integrals in $(\ref{addterm})$. 
But
they have opposite signs in $(\ref{declogpl})$ (compare $(\ref{declogpl})$  with $(\ref{declog})$)
and hence their contributions cancel each other totally. This means that $(\ref{etaexpect02})$ does
really coincide with $(\ref{expecteta})$ in the limit $x+y+z+t \rightarrow \infty$,
$Q<1$ (which means conditions $(\ref{exponcond})$ and $(\ref{condeta})$ to be valid).

To find an asymptotic behaviour of $<\eta ^2>$ at large $x+y+z+t$ and $Q=$
$(z+t)/(x+y)<1$ we put  relation $(\ref{wplus01})$  in
 $(\ref{squareta})$  and consider  the limit $a \rightarrow 1$ ($a>1$).
Like in calculating $<\eta>$ the dominant contributions to the integrals in $(\ref{squareta})$ come
from
the two regions $0\leq \phi \leq \beta$, $2 \pi -\beta \leq \phi \leq 2 \pi$ ($\beta \ll 1$). 
These contributions
are complex conjugate quantities. Hence we may consider the real part of the integral over $\phi$ in
the limits 0 and $\pi$. Since the integrals are convergent very rapidly we may use the limits 0 and
$\infty$ neglecting exponentially small corrections. To fulfill this program we have to
establish the
behaviour of the function
\begin{equation} 
L_1=\int_0^{\infty} \frac{\alpha e^{-\alpha}}{1- \theta e^{-\alpha}}d \alpha 
\label{defl1}
\end{equation}
at small $\phi$ where $\theta =e^{-i \phi}$ in $(\ref{defl1})$. The other functions of $\phi$ in 
$(\ref{squareta})$ can be expressed through $L_1$. Indeed, it is obvious that  
\begin{equation}
L_2=\int_0^{\infty} \frac{\alpha e^{-2 \alpha}}{1- \theta e^{-\alpha}}d \alpha =
\frac{L_1}{\theta}-\frac{1}{\theta}\;.
\label{defl2}   
\end{equation}
Applying the new variable $t=e^{-\alpha}$ and integrating several times by parts we get
\begin{eqnarray}
\nonumber
L_1=-\int_0^{1} \frac{\ln t}{1- t \theta }d t\\
\nonumber
=\frac{\ln t}{\theta} \ln (1-t\theta )\Big |^1_0-
\int_0^{1} \frac{dt}{t \theta} \ln (1- t \theta )=\\
\nonumber
\frac{1}{t \theta ^2}[(1- t \theta ) \ln (1- t \theta )+ t \theta ] \Big |^1_0\\
\nonumber
+\int_0^{1} \frac{dt}{t^2 \theta ^2} [(1- t \theta ) \ln (1- t \theta )+t \theta]=\\
\nonumber
\frac{1}{ \theta ^2}[(1-  \theta ) \ln (1-  \theta )+  \theta ]\\
\nonumber  
-\frac{1}{t^2 \theta ^3} \Bigl [\frac{(1- t \theta )^2}{2} \ln (1- t \theta )-
\frac{3}{4} (1- t \theta )^2- t \theta +\frac{3}{4} \Bigr ] \Big |^1_0 \\
\nonumber
-\frac{2}{\theta ^3}\int_0^{1} 
\Bigl [\frac{(1- t \theta )^2}{2} \ln (1- t \theta )-
\frac{3}{4} (1- t \theta )^2- t \theta +\frac{3}{4} \Bigr ] \frac{dt}{t^3} = \\L_1^a+L_1^b+L_1^c.
\label{labc1} 
\end{eqnarray}
It is easy to see that at small $\phi$
\begin{eqnarray}
\nonumber
L_1^a=
\frac{1}{ \theta ^2}[(1-  \theta ) \ln (1-  \theta )+  \theta ] \\
\nonumber
\approx
1+i \phi+(i \phi-\frac{3}{2} \phi^2)(\ln \phi +i \frac{\pi}{2})+...\;,\\
\nonumber
L_1^b=-\frac{1}{ \theta ^3} \Bigl [\frac{(1-  \theta )^2}{2} \ln (1-  \theta )-
\frac{3}{4} (1-  \theta )^2-  \theta +\frac{3}{4} \Bigr ] \\ \approx
\frac{1}{4}-\frac{i}{4} \phi +\frac{5}{8} \phi ^2+\frac{\phi ^2}{2}[\ln \phi +i\frac{\pi}{2}]+...
\label{lab1}   
\end{eqnarray}
and hence the derivatives $\partial L_1^a/\partial \phi$, $\partial ^2 L_1^a/\partial \phi ^2$,
$\partial ^2 L_1^b/\partial \phi ^2$ do not exist at $\phi =0$ since expressions $(\ref{lab1})$
 contain $\ln \phi$. One may check that $\partial L_1^c/\partial \phi$, $\partial ^2 L_1^c/\partial
\phi ^2$ are finite at $\phi =0$ which means that $L_1^c$ may contain terms $\phi ^n \ln \phi$
with $n \geq 3$ only. Differentiating $L_1^c$ over $\phi$ it can be easily obtained that
\begin{eqnarray}
\nonumber
L_1^c=\\
\nonumber
-\frac{2}{\theta ^3}\int_0^{1}
\Bigl [\frac{(1- t \theta )^2}{2} \ln (1- t \theta )-
\frac{3}{4} (1- t \theta )^2- t \theta +\frac{3}{4} \Bigr ] \frac{dt}{t^3} \\ \approx  
-\frac{5}{4}+\zeta(2)+i\phi[-\frac{7}{4}+\zeta(2)]+\frac{\phi ^2}{2}[-\frac{3}{4}+\zeta(2)]+...
\label{lc1}
\end{eqnarray}
where $\zeta(2)=\pi ^2/6$ and $\zeta(z)$ denotes the Riemann $\zeta$-function. 
Putting $(\ref{lab1})$
and $(\ref{lc1})$ into $(\ref{labc1})$ we get 
\begin{eqnarray}
\nonumber
L_1= \zeta(2)+i\phi[\zeta(2)-1]+\frac{\phi ^2}{2}[\zeta(2)+\frac{1}{2}]\\
+(i \phi- \phi^2)(\ln \phi +i \frac{\pi}{2})+...\;\;.
\label{l01}
\end{eqnarray}
The term $\epsilon I_1(2\sqrt{(x+y)(z+t)})$ in $(\ref{squareta})$ is exponentially small owing to 
$(\ref{asympinz})$. Decomposing $e^{\pm i m \phi}$ in $(\ref{squareta})$ into power series, putting
them and $(\ref{l01})$, $(\ref{defl2})$, $(\ref{wplus01})$ into $(\ref{squareta})$ and integrating
over $\phi$ from 0 to $\infty$ one gets
\begin{eqnarray}
\nonumber
<\eta ^2>=(x-y+t-z)^2(\frac{1}{v^2}+\frac{3u}{v^4})+\frac{u}{v^2}+\frac{3u^2}{v^4}\\
-\frac{2}{v^2}-\frac{4}{v^3}[(x-y)^2-(t-z)^2]
\label{asymeta2}
\end{eqnarray}
with $u=x+y+z+t$ and $v=x+y-z-t$.    
Substitution of $(\ref{asymeta2})$ and $(\ref{etaxyztinf})$ into $(\ref{dispeta})$ gives the final
formula $(\ref{varsigma})$ for the variance $\delta \eta ^2$ if conditions $(\ref{exponcond})$
and
$(\ref{condeta})$ are valid.  
%%%\newpage


\begin{thebibliography}{4}
\bibitem{PDG} C.~Caso, G~Conforto, A.~Gurtu et al. (Particle Data Group)
{\em  Eur. Phys. J.} {\bf 3} (1998) 1.
\bibitem{Hudson} D.~J.~Hudson, "Statistics. Lectures on Elementary Statistics and Probability",
Geneva, 1964.
\bibitem{Martin} B.~R.~Martin, Statistics for Physicists, Academic Press, London and
New York, 1971.
\bibitem{Whitt} E.~T.~Whittaker, G.~N.~Watson, "A Course of Modern Analysis",   
Cambridge, 1927.
\end{thebibliography}
\end{document}